%% file: main.tex
\documentclass[conference]{IEEEtran}

\usepackage{listings}
\usepackage{subcaption}
\usepackage{svg}
\usepackage{algorithm}
\usepackage{algpseudocode}

\usepackage{amsmath,amssymb,amsfonts}
\usepackage{mathtools}
\usepackage{xcolor}
\usepackage[hidelinks]{hyperref}
\usepackage{enumitem}
\usepackage{booktabs}
\usepackage{cite}
\usepackage{textcomp}
\usepackage{graphicx}
\usepackage{float}
\usepackage{mdframed}
\usepackage{xspace}
\usepackage{comment}
\usepackage{bbm}
\newfloat{cascade}{hbtp}{lop}
\floatname{cascade}{Cascade}

\lstdefinestyle{compactPython}{
  language=Python,
  basicstyle=\footnotesize\ttfamily,
  keywordstyle=\color{blue},
  commentstyle=\color{gray},
  frame=single,
  breaklines=true,
  captionpos=b,
  aboveskip=0.3em,
  belowskip=0.3em,
  lineskip=0pt,
  xleftmargin=1.5em,
  xrightmargin=0em,
  framexleftmargin=1.5em,
  framexrightmargin=0em,
  numbers=left,
  numberstyle=\tiny\color{gray},
  numbersep=1em,
  showstringspaces=false,
  linewidth=\columnwidth,
  escapeinside={(*@}{@*)}
}

\definecolor{fusionblue}{RGB}{220,235,255}
\definecolor{fusionyellow}{RGB}{255,250,215}

\newcommand{\fgone}[1]{\colorbox{fusionblue}{#1}}
\newcommand{\fgtwo}[1]{\colorbox{fusionyellow}{#1}}

\newcommand{\phase}[1]{\shortintertext{\hspace{1em}\textit{#1}}}

\newcommand{\tensor}[1]{\mathit{#1}}

\newcommand{\ED}{\tensor{ED}}
\newcommand{\EY}{\tensor{EY}}
\newcommand{\HH}{\tensor{HH}}
\newcommand{\HX}{\tensor{HX}}
\newcommand{\LEX}{\tensor{LEX}}
\newcommand{\LY}{\tensor{LY}}
\newcommand{\MEX}{\tensor{MEX}}
\newcommand{\NEX}{\tensor{NEX}}
\newcommand{\NUM}{\tensor{NUM}}
\newcommand{\NY}{\tensor{NY}}
\newcommand{\PX}{\tensor{PX}}
\newcommand{\RX}{\tensor{RX}}
\newcommand{\SQEX}{\tensor{SQEX}}
\newcommand{\SSIXY}{\tensor{S6Y}}
\newcommand{\Tokens}{\tensor{Tokens}}
\newcommand{\TTX}{\tensor{TTX}}
\newcommand{\TT}{\tensor{TT}}
\newcommand{\TX}{\tensor{TX}}
\newcommand{\WB}{\tensor{WB}}
\newcommand{\WC}{\tensor{WC}}
\newcommand{\WConv}{\tensor{WConv}}
\newcommand{\WET}{\tensor{WET}}
\newcommand{\WEX}{\tensor{WEX}}
\newcommand{\WEY}{\tensor{WEY}}

\newcommand{\WRX}{\tensor{WRX}}
\newcommand{\WTTX}{\tensor{WTTX}}
\newcommand{\WTT}{\tensor{WTT}}
\newcommand{\WTX}{\tensor{WTX}}
\newcommand{\WT}{\tensor{WT}}
\newcommand{\TB}{\tensor{TB}}
\newcommand{\TC}{\tensor{TC}}
\newcommand{\TTB}{\tensor{TTB}}
\newcommand{\TTC}{\tensor{TTC}}
\newcommand{\WTB}{\tensor{WTB}}
\newcommand{\WTC}{\tensor{WTC}}
\newcommand{\WTTB}{\tensor{WTTB}}
\newcommand{\WTTC}{\tensor{WTTC}}
\newcommand{\LLY}{\tensor{LLY}}
\newcommand{\PLLY}{\tensor{PLLY}}
\newcommand{\PLLYNUM}{\tensor{PLLYNUM}}
\newcommand{\MLLY}{\tensor{MLLY}}
\newcommand{\SQLLY}{\tensor{SQLLY}}
\newcommand{\NLLY}{\tensor{NLLY}}
\newcommand{\WNLLY}{\tensor{WNLLY}}

\begin{document}

\makeatletter
\newcommand{\ALG@resetline}{\setcounter{ALG@line}{0}}
\makeatother

\pdfpagewidth=8.5in
\pdfpageheight=11in

\pagenumbering{arabic}

\title{Mambalaya: Einsum-Based Fusion Optimizations on State-Space Models}

\author{
\IEEEauthorblockN{Toluwanimi O. Odemuyiwa}
\IEEEauthorblockA{
\textit{University of California, Davis} \\
Davis, CA, USA \\
todemuyiwa@ucdavis.edu}
\and
\IEEEauthorblockN{John D. Owens}
\IEEEauthorblockA{
\textit{University of California, Davis} \\
Davis, CA, USA \\
jowens@ucdavis.edu}
\and
\IEEEauthorblockN{Joel S. Emer}
\IEEEauthorblockA{
\textit{MIT / NVIDIA} \\
Cambridge, MA, USA \\
emer@csail.mit.edu}
\and
\IEEEauthorblockN{Michael Pellauer}
\IEEEauthorblockA{
\textit{NVIDIA} \\
Westford, MA, USA \\
mpellauer@nvidia.com}
}

\maketitle
\thispagestyle{plain}
\pagestyle{plain}

\begin{abstract}
\input{sections/0-abstract}
\end{abstract}

\input{sections/1-intro}

\input{sections/3-motivation}
\input{sections/9-related}
\input{sections/4-einsum_analysis}
\input{sections/5-mamba_fusion}

\input{sections/7-binding}
\input{sections/8-evaluation}
\input{sections/10-conclusion}

\clearpage
\bibliographystyle{IEEEtranDOI}
\bibliography{bibs/mamba}

\clearpage
\appendices
\input{sections/appendix.tex}

\end{document}

%% file: sections/0-abstract.tex
Mamba is an emerging, complex workload with various short-range and long-range dependencies, nonlinearities, and elementwise computations that are unable to run at near-peak speeds on modern hardware. Specifically, Mamba's complex dependency graph makes fusion across its full operator cascade difficult, leaving substantial inter-operator memory traffic on the table.
To address these challenges, we propose Mambalaya, a novel reconfigurable accelerator that leverages fusion to overcome the limitations of Mamba. We use the recently proposed cascade-of-Einsums abstraction to characterize Mamba's full computational structure, then apply the extended Einsum framework to systematically explore inter-Einsum fusion opportunities. This principled approach yields a series of fusion mappings that reduce off-chip inter-Einsum traffic. These mappings are supported by the underlying Mambalaya architecture. Mambalaya achieves a layer performance speedup of 4.9$\times$ for prefill and 1.9$\times$ for generation over state-of-the-art Mamba accelerator, MARCA, and 1.5$\times$ in prefill over a recent, fine-grained fusion Mamba accelerator.

%% file: sections/1-intro.tex
\section{Introduction}\label{sec:intro}

Sequence modeling~\cite{sutskever:2014:SSL}---the ability to generate and predict future data or \emph{tokens} based on an input sequence---has emerged as the key generative AI workload. The Transformer network~\cite{Vaswani:2017:AAN} has emerged as a major front-runner, but concerns over its limited ability to retain context over long sequences of tokens~\cite{Chen:2023:CLEX, Li:2024:ECW, chowdhery2022palm} have led to a renewed interest in State Space Models (SSMs).
Mamba~\cite{Gu:2023:MLT} is an emerging alternative SSM architecture that efficiently supports large contexts for LLMs.
Mamba passes a fixed-size tensor (called the \emph{hidden state})  recurrently between the same layer of subsequent steps. While this state represents a new inter-token data dependency, it is not any more serial than the Transformer's existing auto-regressive dependency and therefore is a good candidate for efficient generation.
Since 2023, Mamba has been widely adopted by a wide variety of companies and frameworks, including Microsoft~\cite{ren:2024:SHS}, NVIDIA~\cite{nvidia:2025:nfa}, Mistral AI~\cite{mistral:2024:CMF}, and IBM~\cite{huggingface:2024:IEH}, among others~\cite{wang:2024:MLD,Lieber:2024:JHT, zuo:2024:FMC, Behrouz:2024:GMT, Hatamizadeh:2024:MVH, Zhu:2024:VME}.
Additionally, other competitive Mamba-inspired models have emerged, including Griffin~\cite{De:2024:GMG}, Mamba-2~\cite{Dao:2024:TSG}, and Gated Delta Networks~\cite{yang:2025:GDN}.

Today, optimized Transformers consistently run at near-peak speeds on modern accelerators: compute-bound for the \emph{prefill} phase, and memory-bound for the \emph{generation} (or \emph{decode}) phase. In contrast, we show that Mamba implementations do not reach close to peak utilization due to the significantly larger dependency graph of operators---especially a notably higher fraction of complex non-GEMM operations than Transformers.
These characteristics make it harder to both optimize the underlying computation and to find straightforward mappings onto existing unmodified accelerators.

Recent work has demonstrated some initial success optimizing Mamba using inter-operation \emph{fusion}.  Intermediates between producer-consumer operators incur expensive DRAM traffic as entire tensors are dumped to main memory (too large to be efficiently filtered by caching). Fusion has the potential to remove this traffic, thus increasing computational intensity. Thus the benefit of fusion is directly related to the number of operator pairs from Mamba's complex dependency graph that can be fused.
MARCA~\cite{Li:MARCA}, the first published custom accelerator for Mamba, identified intra- and inter-operation memory traffic staging as a key optimization tradeoff space. MARCA uses a custom buffer that can dynamically configure itself between the two; however, the inter-operation fusion scenarios it could exploit are limited. Recently Geens et al.~\cite{geens2025finegrainedfusion} report preliminary findings that identifies fusion as the key ``missing piece'' of Mamba acceleration. They are able to extend inter-operation fusion to the SSM itself, but not the normalization, discrete weight generation, result production, or projection portions of Mamba. Unfortunately, as we demonstrate, because of the sheer complexity of Mamba's non-GEMM operators, fusion opportunities have been only incompletely realized in previous work and do not yet achieve peak performance.

In this work, we are the first to demonstrate that the scope of fusion can be expanded \emph{across the entire Mamba computational sequence}, resulting in minimal inter-operator memory traffic and maximum computational intensity. To accomplish this, we leverage Einstein summations, or \emph{Einsums}, an approach that formally and succinctly describes workloads as a connected graph of tensor algebra operations~\cite{Kjolstad:TACO}. Previous work has demonstrated that this formulation enables systematic and tool-based exploration of the space of possible implementations~\cite{Nayak:2023:TDF_micro, timeloop, Kwon:2019:URP, gilbert:2024:LEF}. Einsums allow for a separation of concerns between a workload's algorithm, mapping, binding, and architectural specifications, and have been successfully used to productively architect a range of custom accelerators across a wide range of tensor algebra applications~\cite{Nayak:2024:FML_micro, Odemuyiwa:2023:ASD, Wu:2022:SAA}

We demonstrate that the conception of inter-Einsum fusion in previous work is not sufficient to optimize Mamba. Therefore, in this work we extend the state of the art in Einsum analysis with 3 key additions: (1)~we demonstrate how the SSM hidden state can be expressed as an iterative tensor index; (2)~we demonstrate how inter-Einsum fusion opportunities relate directly to the tensor indices used in fusion pairs; and (3)~we develop a taxonomy of fusion options that demonstrate new opportunities not shown by previous work. Finally, we leverage our novel fusion techniques to propose Mambalaya, a custom accelerator for Mamba with an array specifically designed for its fusion needs. Because of its increased computational intensity, Mambalaya increases performance by 4.9$\times$ for prefill and 1.9$\times$ for generation over the previous state-of-the-art custom accelerator.

Our specific contributions are as follows:
\begin{enumerate}

\item We define a set of inter-Einsum fusion techniques, taxonomizing previous work and expanding the overall map-space of optimization possibilities.

\item We demonstrate that all tensor algebra Mamba can be expressed as Einsums, allowing its inter-operation complexity to be expressed precisely.

\item Leveraging the above, we are the first to demonstrate that fusion is possible across the entire Mamba cascade despite its significantly higher complexity than Transformers, reducing it to a single fusion group.

\end{enumerate}

%% file: sections/3-motivation.tex
\section{Einsum-Based Analysis of Mamba}\label{sec:motivation}
\label{sec:mamba_einsum}

\subsection{Einsum-Based Architecture Design Methodology}\label{ssec:background:dse}
Our analysis is based on the framework of \emph{tensors} and \emph{Einstein summations} or \emph{Einsums}, following the terminology proposed in TeAAL~\cite{Nayak:2023:TDF_micro} and the ExtenDed General Einsums Language (EDGE)~\cite{Odemuyiwa:2024:edge}. This approach has been well-explained in previous work and is well-supported by several tools in the computer architecture community, and so for space considerations we omit a discussion of its exact syntax.

Recently, FuseMax~\cite{Nayak:2024:FML_micro} successfully applied an Einsum-based methodology to discover a new attention implementation.
This was possible as extended Einsums enable a separation of concerns to the design process~\cite{jrk:2013:halide,Dijkstra:1982:RST}.
This in turn provides what we can call an ``Einsum recipe'' to systematically explore the design space, starting with an algorithm and ending with an accelerator.

Specifically, TeAAL provides a hierarchy of specifications~\cite[Figure 7]{Nayak:2023:TDF_micro} for each concern.
At the top of the hierarchy is the extended Einsum cascade, precisely specifying the algorithm.
Section~\ref{sec:mamba_einsum} presents a diagrammatic view of the Mamba cascade.

The next level in the hierarchy is the \emph{mapping space} (traversal order, partitioning strategy, parallelism, etc.).
In this work, we focus on \emph{fusion}.
The final level of the hierarchy is \emph{binding}, which schedules computation specific hardware units at specific points in execution time.
In this work, we show that fusion also has a binding component, as intermediate data resulting from the chosen mapping must fit on-chip.
Overall, fusion keeps data shared between Einsums on-die in order to reduce off-chip memory accesses on that data.
Mambalaya is a novel accelerator that views Mamba as an extended Einsum cascade and enables the proposed fusion mappings through specific bindings.

In this section we formalize Mamba as a cascade of dependent tensor operations expressed as Einsums.  For pedagogical clarity, we represent the cascade graphically.
The Einsum cascade for a typical Transformer layer, as captured previously by Nayak et al.~\cite{Nayak:2024:FML_micro}, has the following notable features: (A) a small number of overall operators (8 per layer), (B) a relative prevalence of GEMM-like operators (6 out of 8), and (C) a relative simplicity of producer-consumer dependencies and short lifetimes of intermediates.

\begin{figure}
    \centering
\includegraphics[width=0.75\columnwidth, trim=10 40 10 45,
    clip]{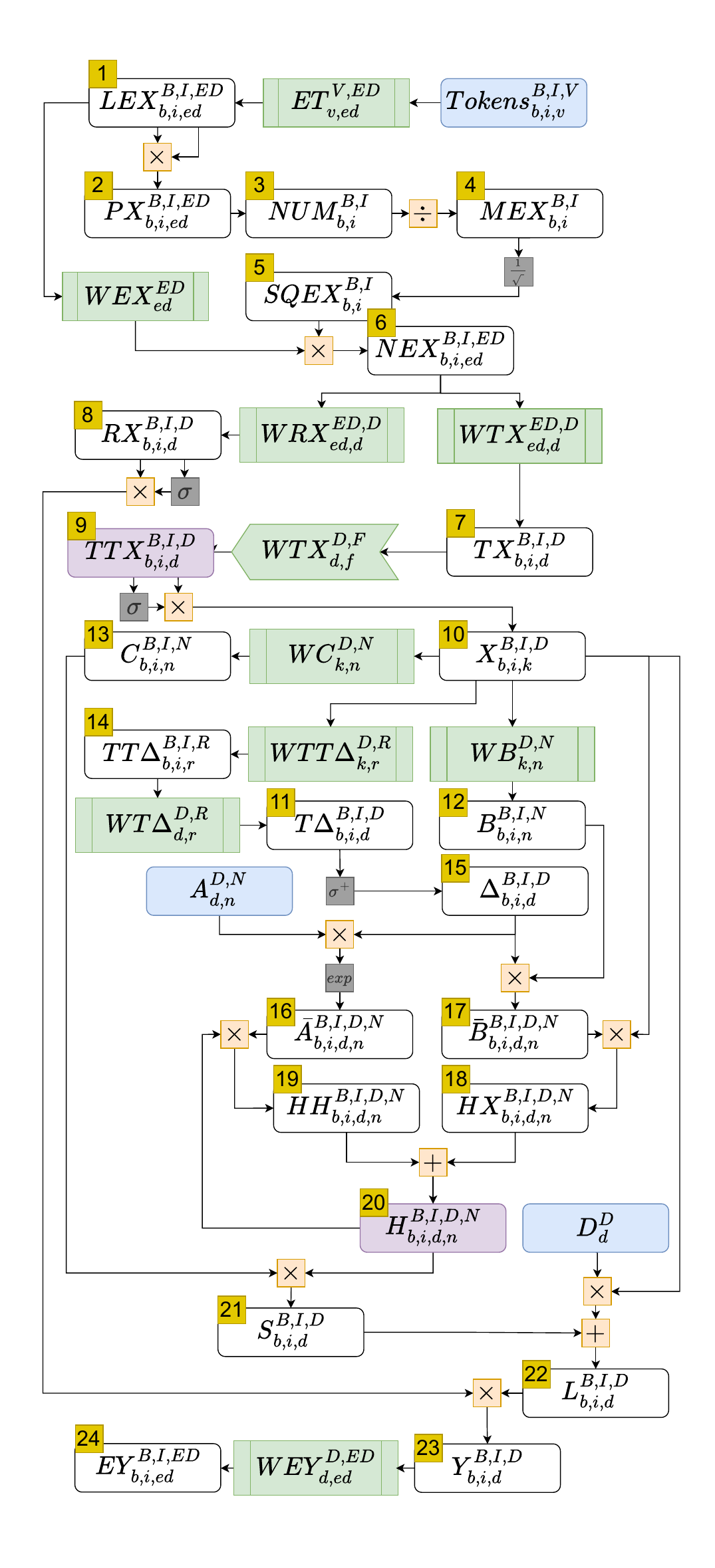}

\caption{Overview of the cascade execution flow for Mamba. Rounded, rectangular boxes are tensors, with the rank names (and shapes) in superscripts, and the corresponding rank variables (tensor indices) in the subscripts~\cite{Odemuyiwa:2024:edge}.
Each Einsum is labeled with a number (yellow box) on its output tensor.
Colors represent the following: (a) blue: input tensor, (b) green: GEMM with a weight tensor, (c) purple: tensor with recurrent accesses (e.g., $H_{i-1}$), (d) light orange: elementwise/broadcast operation, (e) dark grey: unary and nonlinear functions.
    }
    \label{fig:cascade}
\end{figure}

In contrast, Figure~\ref{fig:cascade} shows the cascade for Mamba. It contains 24 distinct tensor operations per layer, with a complex set of dependencies and liveness distances of intermediate values. 7 of those 24 are GEMM-like, but as we show in detail many of them struggle to reach the compute-bound region on modern accelerator hardware. In the rest of this section we explain these equations in detail and quantify the existing challenges and available acceleration opportunity.

\subsection{Structured State-Space Model Component}

Here is the general Einsum cascade for any SSM\@:

\begin{subequations}
\begin{align}\label{eqn:ssm_all}
\HH_{i, n, c} = A_{n} \cdot H_{i-1, n, c} \\
H_{i, n, c}  = B_{i, n, k} \cdot X_{i, k} + \HH_{i, n, c}
\end{align}
\end{subequations}

The $H$ tensor is the \emph{hidden state}; it represents the memory of the system.
On each iteration, it takes as input some scaling of the previous state ($A \cdot H$, stored in the temporary tensor $\HH$) and additively combines this with some scaling of the current input ($B \cdot X$).

Different SSMs vary in how they initialize $A$, $B$, or $C$ to produce different behaviors.
Additionally, prior to Mamba, all SSMs used the same $A$, $B$, $C$ tensors for every token; that is, there was no time-dependent, iterative rank ($i$) on these tensors~\cite{gu:2020:hippo, gu:2021:combining, gu:2022:efficiently, goel:2022:sashimi, gu:2023:hippo, gu:2022:s4d, nguyen:2022:s4nd}.
To change how the model remembers information based on the current input token (``input selectivity''~\cite{Gu:2023:MLT}), Mamba adds the $I$ rank to all three tensors by making them the result of GEMM and non-linear operations on the input tokens (see $\bar{A}$, $\bar{B}$, $\bar{C}$ in Figure~\ref{fig:cascade}).

The input selectivity of Mamba enables the model to decide which inputs to focus on and which to ignore.
In attention-based models, the KV cache contains an uncompressed record of every token ever seen, growing linearly in size as token length increases.

On the other hand, the $H$ state tensor in Mamba maintains a consistent state size (rank $N$), acting as a ``compressed summary'' of all the tokens seen so far and their relative importance.
The larger $N$ is, the less compressed $H$ will be.
In Mamba-1, $N = 16$.
This constant state size enables Mamba to theoretically record an infinite number of tokens.

A Mamba layer consists of more than just the SSM\@: in the prefill phase, Mamba
(1) translates the entire context to the embedding space,
(2) performs normalization to establish numeric stability,
(3) applies causal correlation to the inputs to capture short-range dependencies,
(4) transforms the SSM tensors ($A$, $B$, $C$) into input-dependent and time-varying (via the $I$ rank) tensors,
(5) applies the SSM, returning a final state $H$ tensor for the prefill phase, and
(6) either casts the result back to token space, generating a new token, or passes it to the next
layer.
In the token generation phase, there is only one token to process at a time.
Each token goes through the same steps as the prefill phase, with the state tensor ($H$) initialized to the previous token's state tensor.

\begin{figure*}
    \footnotesize
    \centering
    \subcaptionbox{Overall roofline utilization}{\includegraphics[width=0.3\textwidth]{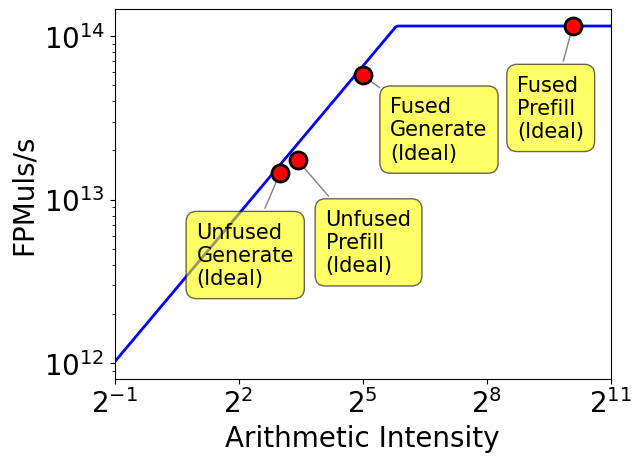}}
    \hspace{0.02\textwidth}
    \subcaptionbox{Prefill: ideal unfused (top) vs fused}{\includegraphics[width=0.32\textwidth]{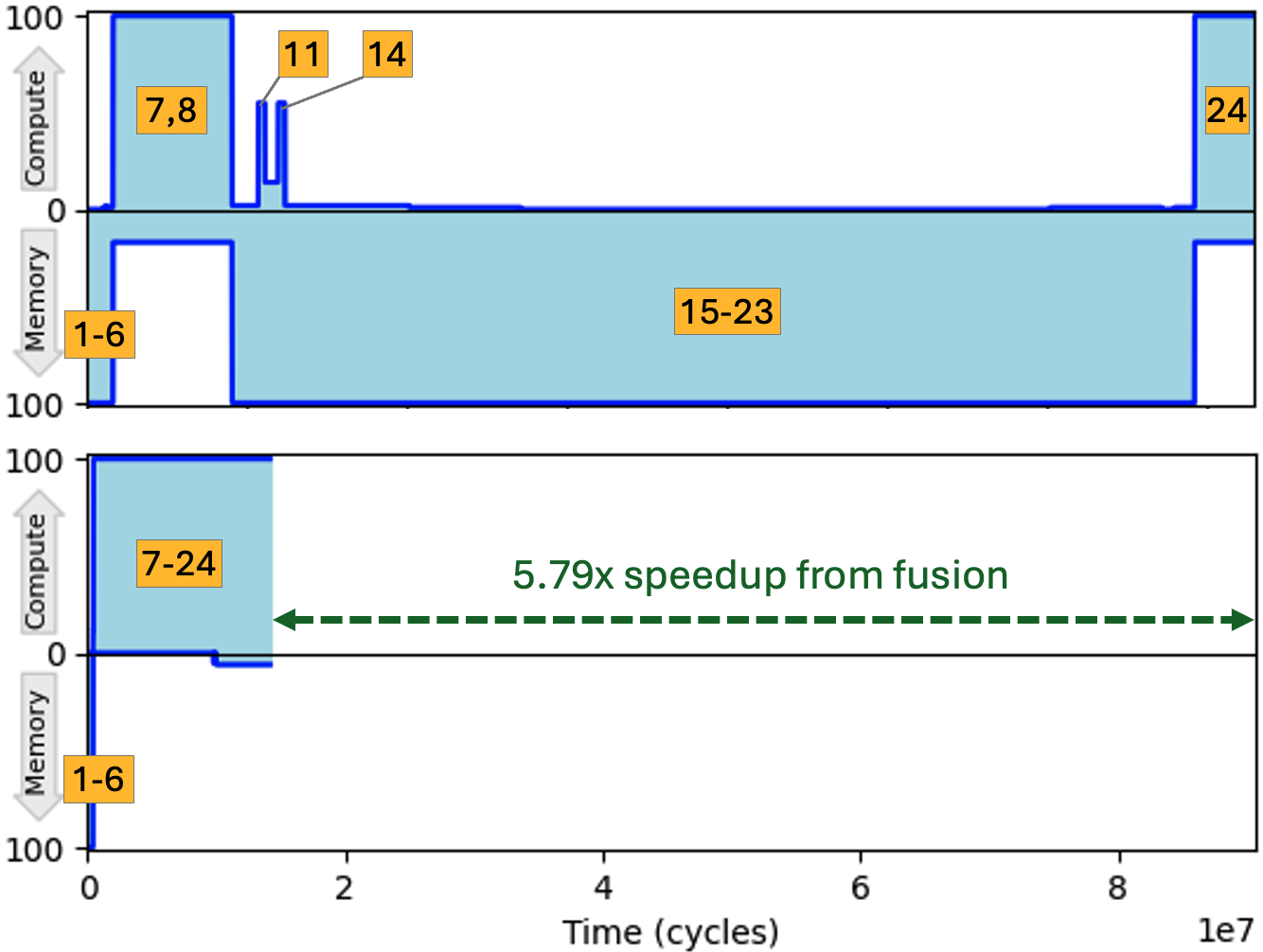}}
    \hspace{0.02\textwidth}
    \subcaptionbox{Generate: ideal unfused (top) vs fused}{\includegraphics[width=0.32\textwidth]{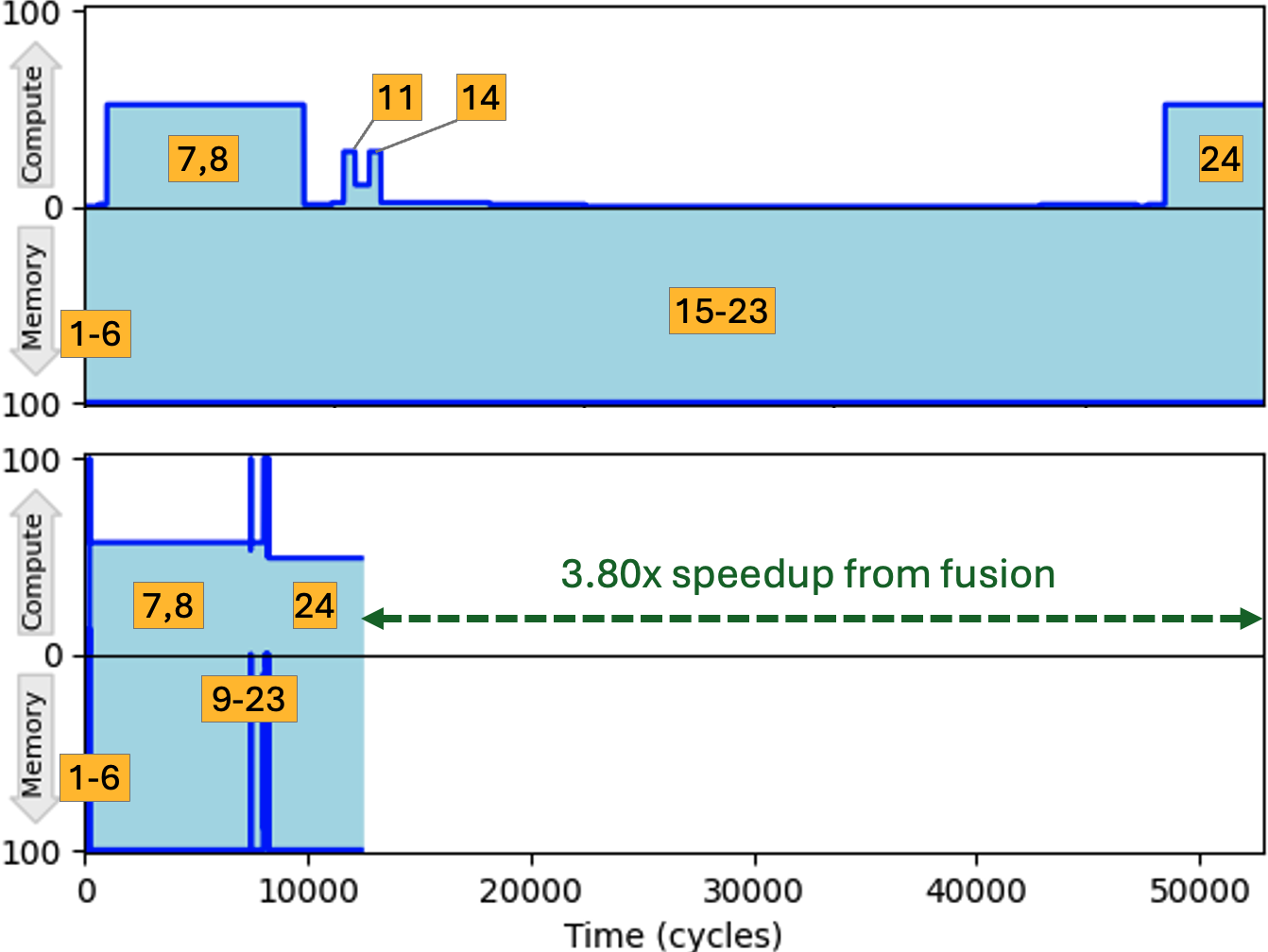}}
    \caption{Overall roofline plot (a) confirms that unfused operations are memory-bound, but does not give insight into Einsums' relative weighting. Plotting detailed roofline utilization over time for a single layer demonstrates that (b) unfused prefill alternates between compute-bound and memory-bound Einsums, whereas (c) decode does not have enough reuse to reach the compute-bound in any Einsum. Phase labels in yellow correspond to the Einsum numbers in Figure \ref{fig:cascade}. Ideal fusion would eliminate all inter-Einsum traffic, resulting in significant increases to intensity, as shown in the bottom figures.}
    \label{fig:roofline}
\end{figure*}

\subsection{Analysis: Challenges and Opportunities}

\begin{table}
\centering
\caption{Traffic breakdowns for a single layer of the best-case Mamba accelerator without fusion (Best Unfused).
}
\label{tab:unfused-traffic}
\footnotesize
\begin{tabular}{lc|lc}
\toprule
\textbf{Traffic Type} & \textbf{Percentage} & \textbf{Traffic Type} & \textbf{Percentage}\\
\midrule
Read Traffic          & 47.3\% & Inter-Einsum  & 99.1\% \\
Write Traffic         & 52.7\% & Intra-Einsum  & 0.9\% \\
\bottomrule
\end{tabular}
\vspace{-10pt}
\end{table}

The cascade in Figure~\ref{fig:cascade} contains two sources of parallelism: intra-Einsum (where available parallelism is limited by operand tensor dimensions) and inter-Einsum (where available parallelism is limited by dependency edges). To start, let us consider an accelerator that focuses solely on intra-Einsum parallelism.
In this implementation, we execute each Einsum in the cascade sequentially, producing the complete output tensor before moving to the next Einsum.
Each input tensor is read from main memory and each output tensor is written back to main memory in completeness.
We assume that the accelerator has sufficient on-chip buffering to achieve the algorithmic minimum DRAM access traffic---where each tensor is accessed once for the specified Einsum without additional spills/fills.
Table~\ref{tab:unfused-traffic} shows the total input/output traffic this implementation would achieve.  We call this the \emph{Best Unfused} implementation and refer to it throughout the paper.
Some of this traffic is for tensors that are used again by subsequent Einsums.
Let us label the traffic for tensors shared across Einsums as \emph{inter-Einsum} traffic, and the traffic for tensors unique to an Einsum as \emph{intra-Einsum} traffic.

Table~\ref{tab:unfused-traffic} gives a detailed breakdown of the traffic for a best-case unfused implementation that has sufficient buffering to achieve perfect data reuse within each Einsum. Figure \ref{fig:roofline}(a) uses roofline analysis \cite{williams:2009:RIV} to demonstrate that even with this optimistic assumption, sequential unfused implementations of this Einsum cascade are fundamentally memory-bound. Beyond this, Figures \ref{fig:roofline}(b) and (c) plot the arithmetic intensity over time across the 24 Einsums.
If all inter-Einsum traffic was eliminated (i.e., ideal fusion) it would result in significant speedups (5.79$\times$ for prefill, and 3.8$\times$ for token generation), as well as energy efficiency gains from the traffic reductions.

All in all, this demonstrates the great potential of fusion: if we could reduce the inter-Einsum traffic to zero it would result in a tremendous efficiency increase.
Of course, this is not straightforward---re-dedicating on-chip buffering to holding inter-Einsum intermediates takes it away from buffering intra-Einsum operands. When analyzing the efficiency of a given Einsum, there are three components to focus on: compute resource utilization, memory accesses, and overall latency.

%% file: sections/9-related.tex
\subsection{Related work}

\subsubsection{Fusion in Tensor Algebra Workloads}
\begin{table*}
\centering
\small
\renewcommand{\arraystretch}{1.1}
\setlength{\tabcolsep}{4pt}
\begin{tabular}{lcccccccc}
\toprule
\textbf{Work} &
\textbf{RI} &
\textbf{RSb} &
\textbf{RSp} &
\textbf{RD} &
\textbf{Stitching} &
\textbf{Minimum ITF} &
\textbf{Workloads} &
\textbf{Target Architecture} \\
\midrule

\textbf{XLA-like}
~\cite{Shen:2021:Nimble,Zhu:2021:DISC,Sabne:2020:XLA,Snider:2023:OFX} &

\checkmark &  &  &  & RI & Unit & DL & CPU, GPU  \\
\textbf{TVM}~\cite{Chen:2018:TVM}, \textbf{AStitch}~\cite{Zheng:2022:AStitch} & \checkmark &  & \checkmark* &  & RI & Unit, Tile & DL & CPU, GPU, TPU \\

\textbf{PyTorch-Like}~\cite{Zheng:2023:BLD,NVIDIA:2025:TensorRT,Ansel:2024:Pytorch}

& \checkmark & \checkmark & \checkmark* &  &
\shortstack[l]{RI+RSb+RSp} & Unit, Tile & DL & CPU, GPU\\

\textbf{APOLLO} \cite{Zhao:2022:APOLLO} & \checkmark & \checkmark & \checkmark & \checkmark* &
\shortstack[l]{RI+RSb+RSp} & Unit, Tile & DL & GPU, Huawei Ascend \\

\textbf{CNN DSAs}~\cite{waeijen:2021:CMF,cai:2023:ISD} & \checkmark &  & \checkmark &  &
\shortstack[l]{RI+RSp, Recomp.} & Tile & CNNs & DSA \\

\textbf{TileFlow} \cite{Zheng:2023:TFF} & \checkmark & \checkmark & \checkmark &  &
\shortstack[l]{RI+RSb+RSp, Recomp.} & Tile & DL & DSA \\

\textbf{LoopTree} \cite{gilbert:2024:LEF} & \checkmark & \checkmark & \checkmark & \checkmark &
\shortstack[l]{RI, Recomp.} & Tile & DL,TA& DSA \\

\textbf{MARCA} \cite{Li:MARCA} & \checkmark &  &  &  & RI & Tile & Mamba-1 & DSA\\

\textbf{Geens et al.} \cite{geens2025finegrainedfusion} & \checkmark &  &  &  & RI & Unit, Tile & Mamba-1 & DSA \\

\textbf{This Work (Mambalaya)} &
\checkmark & \checkmark & \checkmark & \checkmark &
All combos &
Unit, Tile (RD) &
\shortstack[l]{Mamba-1/2, TA+} &
DSA \\
\bottomrule
\end{tabular}
\caption{Fusion in related works. ITF - Intermediate Tensor Footprint, TA - Tensor Algebra, DSA - Domain specific accelerator, DL - Deep Learning. ``*'' indicates the fusion type is supported with limitations. Our fusion taxonomy supports Mamba-1, Mamba-2, and any workload expressible as an EDGE~\cite{Odemuyiwa:2024:edge, Nayak:2023:TDF_micro} cascade (TA+).
}
\label{tab:relatedworks}
\end{table*}

Fusion, in the context of tensor algebra, has mostly been relegated to neural network accelerators, where various works have focused on inter-layer fusion~\cite{gilbert:2024:LEF, Zheng:2023:TFF, Kwon:2019:URP, flat, waeijen:2021:CMF, cai:2023:ISD, cai:2023:OFF} to remove the need for sending entire layer outputs off-chip.
These all fall into combinations of RI, RSb, and RSp fusion, our proposed fusion classifications (Section~\ref{sec:fusion}).
Table~\ref{tab:relatedworks} shows the space of fusion in prior work.

Within this space, TileFlow~\cite{Zheng:2023:TFF} exposed a 3D design space that can explore loop order, binding, and tiling options for fusion between layers.
Looptree~\cite{gilbert:2024:LEF} allows for the modeling of an even larger design space that considers recomputation and more tiling options, although it does not search the space.
FuseMax~\cite{Nayak:2024:FML_micro} introduces pass analysis to transform a cascade into one that is more amenable to RI, RSb, and RSp fusion.

Both MARCA~\cite{Li:MARCA} and Geens et al.~\cite{geens2025finegrainedfusion} focus on the SSM portion for fusion (Section~\ref{sec:eval}).
Due to the extended Einsum framework and our clearly identified taxonomy, we are able to apply fusion to not only the SSM, but the \emph{entire} workload.

%% file: sections/4-einsum_analysis.tex
\section{Fusing Extended Einsums}\label{sec:fusion}
\subsection{Einsum-based Definition of Fusion}\label{sec:fusion_gen}
\begin{figure*}
  \centering

  \begin{subfigure}[t]{0.23\textwidth}
    \centering
    \includegraphics[width=\linewidth,
    trim=0 10 10 0,
    clip
    ]{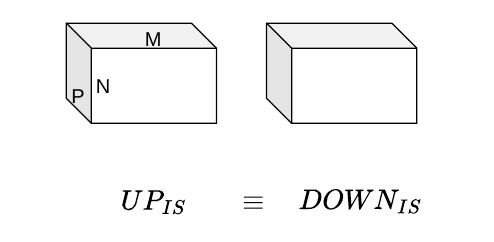}
    \caption{RI Fusion}
    \label{fig:fusion-ri}
  \end{subfigure}
  \hfill

  \begin{subfigure}[t]{0.23\textwidth}
    \centering
    \includegraphics[width=\linewidth,
    trim=0 10 10 0,
    clip
    ]{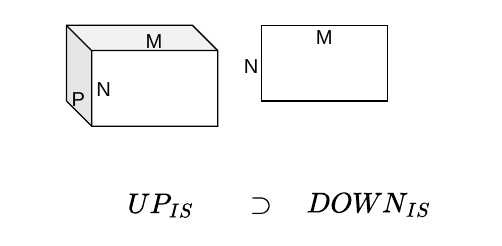}
    \caption{RSb Fusion}
    \label{fig:fusion-rsb}
  \end{subfigure}
  \hfill

  \begin{subfigure}[t]{0.23\textwidth}
    \centering
\includegraphics[width=\linewidth,
    trim=0 10 10 0,
    clip
    ]{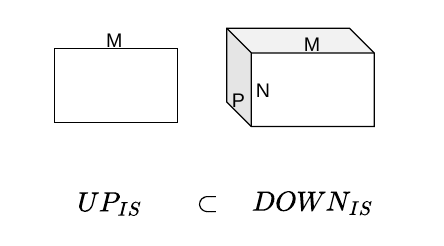}
    \caption{RSp Fusion}
    \label{fig:fusion-rsp}
  \end{subfigure}
  \hfill

  \begin{subfigure}[t]{0.23\textwidth}
    \centering
    \includegraphics[width=\linewidth,
    trim=0 10 10 0,
    clip
    ]{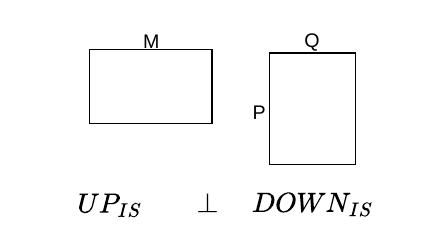}
    \caption{RD Fusion}
    \label{fig:fusion-rd}
  \end{subfigure}

  \caption{Examples of how each fusion class transforms an upstream iteration space to a downstream iteration space.}
  \label{fig:fusion-classes-iteration-space}
  \vspace{-10pt}
\end{figure*}
Fusion places constraints on three levels of the separation-of-concerns pyramid proposed in TeAAL~\cite[Figure 7]{Nayak:2023:TDF_micro}: (1)~Einsum level, (2)~mapping level and (3)~binding and architecture.
Fusion can occur at any level of the memory hierarchy, with fusion at one level minimizing inter-Einsum traffic to that level's backing store (e.g., register to on-chip buffer or on-chip buffer to global memory).
We walk through the requirements at each level of the pyramid:

\par\textbf{{Einsum level}}
Each Einsum has its own iteration space.
Fusion is possible between two Einsums only if the upstream Einsum has an output tensor that is an input to the downstream Einsum.
We call this shared tensor the \emph{intermediate tensor}.

\par{\textbf{Mapping level.}}
An unfused implementation traverses the entire upstream iteration space ($IS_{up}$) before traversing the downstream iteration space ($IS_{dwn})$.
Fusing these two Einsums requires a combined iteration space composed of their respective iteration spaces.
Then, a fused implementation traverses the combined iteration space, resulting in a traversal that alternates between a portion of the upstream and a portion of the downstream.

\par{\textbf{Binding level.}}
Finally, fusion requires a binding such that, as execution traverses the iteration space, there are no spills or fills of the intermediate shared tensor to its backing store (usually global memory).
Note that with this additional requirement, a mapping that completes traversal of the upstream iteration space, \emph{then} iterates over the downstream \emph{could} be fused if and only if there is on-chip storage large enough to prevent spills and fills of the shared intermediate tensor.
Partial fusion may occur if the intermediate is overbooked~\cite{Xue:2023:TAS}.

\subsection{Challenges}
Given a cascade, a successful fusion strategy must find a mapping and binding that enables communicating between producer Einsums and consumer Einsums without accessing the backing store.
However, this goal encounters several challenges associated with successful fusion.
On-chip inter-Einsum storage reduces the available space for intra-Einsum storage.
Additionally, a mapping that enables fusion to occur may also constrain the dataflow of the upstream or downstream Einsum such that there is poor locality in non-intermediate tensors.
This effect can propagate throughout the entire cascade.

Finally, finding a good fusion strategy, and knowing how to fuse, becomes difficult as cascades grow in complexity.
At the Einsum level, we must find solutions for the following challenges:
(A)~A producer Einsum with multiple consumer/downstream Einsums, (B)~A consumer/downstream Einsum with multiple producer/upstream Einsums, and (C)~Generational ranks with unit sizes (e.g., $H_{i+1} = H_i$) and non-unit step sizes (e.g., the $\TX$ to $\TTX$ Einsum in Figure~\ref{fig:cascade}).

\subsection{Mapping : Fusing a Pair of Einsums}\label{sec:pairwise-fusion}

We first classify fusion between a pair of Einsums.
\emph{The fusion opportunity is dependent on the relationship between the upstream Einsum ($IS_{up}$) and the downstream Einsum ($IS_{dwn}$).}
Fusion should minimize the intermediate footprint in order to maximize buffer space for \emph{intra-}Einsum reuse.
Our classification guarantees the minimum-possible intermediate tensor footprint (ITF) of one.
Regardless of the workload, all fusion opportunities between two Einsums with an output-input tensor relationship fall into the following four categories:

\subsubsection{Rank Isomorphic Fusion (RI)}
\begin{figure}
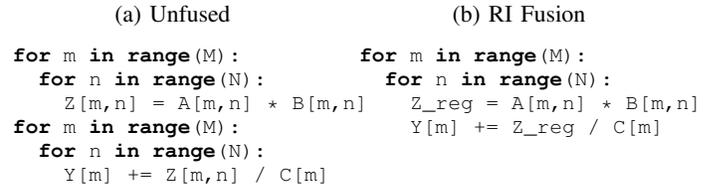

  \centering
  \begin{subfigure}[t]{0.48\columnwidth}
    \centering
    \caption{Unfused}
    \begin{lstlisting}[language=Python,basicstyle=\ttfamily\footnotesize]
for m in range(M):
  for n in range(N):
    Z[m,n] = A[m,n] * B[m,n]
for m in range(M):
  for n in range(N):
    Y[m] += Z[m,n] / C[m]
    \end{lstlisting}
  \end{subfigure}
  \hfill
  \begin{subfigure}[t]{0.48\columnwidth}
    \centering
    \caption{RI Fusion}
    \begin{lstlisting}[language=Python,basicstyle=\ttfamily\footnotesize]
for m in range(M):
  for n in range(N):
    Z_reg = A[m,n] * B[m,n]
    Y[m] += Z_reg / C[m]
    \end{lstlisting}
  \end{subfigure}

  \caption{Unfused to RI fusion. Both Einsums share the same iteration space.}
  \label{fig:unfused-ri-fused}
\end{figure}

If the two Einsums have identical iteration spaces ($IS_{up} \equiv IS_{dwn}$), we can apply \emph{rank isomorphic fusion}.
RI applies an output-stationary dataflow on the upstream Einsum, and an input-stationary dataflow on the downstream.
This ensures a minimum ITF of one on-chip, as an element in the upstream can pass directly to its consumer with a liveness distance of one.
This does not preclude partitioning: a tile of data may pass from one to the other, and the consumer may even choose to read the buffer in reverse order (which prior work does not allow~\cite{Ansel:2024:Pytorch}).
Terms such as elementwise-elementwise, elementwise-reduce, reduce-elementwise used as distinct fusion types in prior work~\cite{Shen:2021:Nimble,Zhu:2021:DISC,Sabne:2020:XLA,Snider:2023:OFX, Chen:2018:TVM, Zheng:2022:AStitch, Zheng:2023:BLD, Ansel:2024:Pytorch} all fall under RI fusion.
Figure~\ref{fig:unfused-ri-fused} shows RI fusion on an elementwise-reduce cascade pattern.

\subsubsection{Rank Subsetted Fusion (RSb)}
\begin{figure}
  \centering
  \begin{subfigure}[t]{0.48\columnwidth}
    \centering
    \caption{Unfused}
    \begin{lstlisting}[language=Python,basicstyle=\ttfamily\footnotesize]
for m in range(M):
  for k in range(K):
    Z[m] += A[m,k] * B[k]
for m in range(M):
  Y[m] = Z[m] / C[m]
    \end{lstlisting}
  \end{subfigure}
  \hfill
  \begin{subfigure}[t]{0.48\columnwidth}
    \centering
    \caption{RSb Fusion}
    \begin{lstlisting}[language=Python,basicstyle=\ttfamily\footnotesize]
for m in range(M):
  for k in range(K):
    Z_reg += A[m,k] * B[k]
    Y[m] = Z_reg / C[m]
    \end{lstlisting}
  \end{subfigure}

  \caption{Unfused to RSb fusion. The upstream Einsum contains a rank ($K$) absent from the downstream.}
  \label{fig:unfused-rsub-fusion}
\end{figure}

In rank subsetted fusion (RSb), the iteration space of the upstream must be a proper superset of the downstream iteration space ($IS_{up} \supset IS_{dwn}$).
This occurs in cascades where the output of the upstream is the result of a reduction.
With an upstream, output stationary and downstream, input stationary dataflow, RSb guarantees a minimum ITF of one.

Figure~\ref{fig:unfused-rsub-fusion} shows an RSb fusion mapping on a matrix-vector computation followed by an elementwise operation.
Note that if we change the mapping to be $KM$ stationary instead of $MK$ stationary, fusion will not be guaranteed: an entire $M$-fiber of $Z$, and all $K$ of its partial products, is produced before the downstream can consume it.

Patterns in prior work that fall into this category include: reduce-elementwise, reduce-reduce, and matrix-matrix multiplication to elementwise~\cite{Zhao:2022:APOLLO}.
Note that prior work differentiates between fusing memory-intensive operations and compute-intensive operations~\cite{Zheng:2022:AStitch, Zheng:2023:BLD}.
With our taxonomy, any Einsum pair that fits the fusion type criteria is fusible.

\subsubsection{Rank Supersetted Fusion (Rsp)}

\begin{figure}
  \centering
  \begin{subfigure}[t]{0.48\columnwidth}
    \centering
    \caption{Unfused}
    \begin{lstlisting}[language=Python,basicstyle=\ttfamily\footnotesize]
for m in range(M):
  for n in range(N):
    Z[m,n] = A[m,n] * B[n]
for m in range(M):
  for n in range(N):
    for p in range(P):
      Y[m,p] += Z[m,n] * C[n,p]
    \end{lstlisting}
  \end{subfigure}
  \hfill
  \begin{subfigure}[t]{0.48\columnwidth}
    \centering
    \caption{RSp Fusion}
    \begin{lstlisting}[language=Python,basicstyle=\ttfamily\footnotesize]
for m in range(M):
  for n in range(N):
    Z_reg = A[m,n] * B[n]
    for p in range(P):
      Y[m,p] +=
          Z_reg * C[n,p]
    \end{lstlisting}
  \end{subfigure}

  \caption{Unfused to RSp fusion. The downstream Einsum contains a rank ($P$) absent from the upstream.}
  \label{fig:unfused-rsp-fusion}
\end{figure}

In rank supersetted fusion (RSp), the iteration space of the upstream must be a proper subset of the downstream iteration space ($IS_{up} \subset IS_{dwn}$).
This occurs in cases when the downstream Einsum \emph{broadcasts} a rank in the upstream Einsum.
Maintaining an upstream, output stationary and downstream, input stationary dataflow ensures a minimum ITF of one.

Figure~\ref{fig:unfused-rsp-fusion} shows RSp fusion on a broadcast Einsum followed by matrix-matrix multiplication. Since ranks $M$ and $N$ are shared, the mapping \emph{must} be $MN$ or $NM$-stationary.

Patterns in prior work include elementwise-broadcast, elementwise-convolution, convolution-convolution (without recomputation)~\cite{Zhao:2022:APOLLO, cai:2023:OFF, gilbert:2024:LEF}.

\subsubsection{Rank Disjointed Fusion (RD)}
\begin{figure}
  \centering
  \begin{subfigure}[t]{0.48\columnwidth}
    \centering
    \caption{Unfused}
    \begin{lstlisting}[language=Python,basicstyle=\ttfamily\footnotesize]
for m in range(M):
  for n in range(N):
    for k in range(K):
      Z[m,n] +=
          A[m,k] * B[k,n]
for m in range(M):
  for n in range(N):
    for p in range(P):
      Y[m,p] +=
          Z[m,n] * C[n,p]
    \end{lstlisting}
  \end{subfigure}
  \hfill
  \begin{subfigure}[t]{0.48\columnwidth}
    \centering
    \caption{RD Fusion}
    \begin{lstlisting}[language=Python,basicstyle=\ttfamily\footnotesize]
for m in range(M):
  for n in range(N):
    for k in range(K):
      Z_reg +=
         A[m,k] * B[k,n]
    for p in range(P):
      Y[m,p] +=
          Z_reg * C[n,p]
    \end{lstlisting}
  \end{subfigure}

  \caption{Unfused to RD fusion. Each Einsum has a rank absent from the other ($K$, $P$).}
  \label{fig:unfused-rd-fusion}
\end{figure}

In rank disjointed fusion (RD), the iteration space of the upstream is neither equal to, a subset of, nor a superset of the downstream iteration space ($IS_{up} \perp IS_{dwn}$).
The upstream iteration space contains one or more ranks absent from the downstream, while the downstream iteration space contains one or more ranks absent from the upstream.
Both a reduction and a broadcast are occurring on the intermediate.
By enforcing an upstream output stationary/downstream input stationary dataflow, RD guarantees a minimum ITF of one.

Figure~\ref{fig:unfused-rd-fusion} applies RD fusion to two back-to-back matrix-multiplication Einsums.
In this case, the mapping \emph{must} be $MN$ or $NM$-stationary, to be upstream output stationary/downstream input stationary.

Patterns in prior work include reduce-broadcast and matmul-matmul~\cite{Zhao:2022:APOLLO, gilbert:2024:LEF}.
Prior work often creates \emph{tiles} of data in this scenario, but our dataflow constraint guarantees a buffer size of one is sufficient.

\noindent
In each of the above cases, partitioning is an orthogonal concern.
Rather than passing unit-sized intermediates between Einsums, a mapping can choose to pass tile-sized intermediates.
However, care must be taken to maintain upstream output-stationarity and downstream input-stationarity.

Figure~\ref{fig:fusion-classes-iteration-space} depicts the relationships between the upstream and downstream iteration spaces.
A fused pair of Einsums forms a \emph{fusion group}.

\subsection{Mapping: Stitching Fusion Groups}
The above taxonomy is simple, yet succinctly covers any fusion strategy between a pair of Einsums.
However, most interesting workloads today are not 2-Einsum Cascades.
\emph{Stitching} merges an Einsum (or fusion group) to an already existing fusion group.
Our proposed stitching strategy greedily forms a fusion group of multiple Einsums whose intermediate tensors stay on-chip.
To fuse larger cascades, we begin by fusing two Einsums in the cascade.
We then greedily check if each subsequent Einsum in the cascade is fusible with the previous fusion group.
A fusion group ends when the last Einsum in the group \emph{must} access the backing store for its output tensor.
This may happen because (A)~the tensor is not an input to any downstream Einsums, (B)~the algorithm \emph{requires} access to the backing store (e.g., the $\LEX$ tensor in Figure~\ref{fig:cascade}), or (C)~there is not enough on-chip storage to keep the tensor on-chip for as long as required.
The second case occurs when an intermediate tensor is part of a two- or three-pass cascade (see FuseMax~\cite{Nayak:2024:FML_micro}),
while the last case may occur when an intermediate tensor becomes an input after execution completes several other Einsums (e.g., the $X$ tensor in Figure~\ref{fig:cascade}).

\subsubsection{Our Greedy Fusion Recipe}
\begin{algorithm}
  \caption{Pseudo-code for Greedy Stitching.}
  \label{alg:greedy-fusion}
  \begin{algorithmic}[1]

    \Require Cascade $C = [E_0, \dots, E_{N-1}]$ \label{alg:req-cascade}
    \Ensure List of fusion groups $\mathcal{F}$ \label{alg:ensure-fgroups}

    \State Initialize empty fusion group $G$ \label{alg:init-G}
    \State $G \gets G \cup \{E_0, E_1\}$ \label{alg:add-first-two}
    \State $IS_{\text{up}} \gets \textsc{IterSpace}(E_0)$ \label{alg:get-rup0}
    \State $IS_{\text{dwn}} \gets \textsc{IterSpace}(E_1)$ \label{alg:get-rdwn1}
    \State $I_{\text{prev}} \gets IS_{\text{up}} \cap IS_{\text{dwn}}$ \label{alg:get-iprev}
    \State $IS_{\text{up}} \gets IS_{\text{dwn}}$ \label{alg:update-rup}

    \For{$i = 2$ to $N-1$} \label{alg:loop-start}
      \State $R_{\text{dwn}} \gets \textsc{IterSpace}(E_i)$ \label{alg:get-rdwn-i}
      \State $I_{\text{curr}} \gets IS_{\text{up}} \cap IS_{\text{dwn}}$ \label{alg:get-icurr}

      \If{
        $\textsc{Subset}(I_{\text{prev}}, I_{\text{curr}})$ \label{alg:cond-subset} \\
        \quad or $\textsc{Superset}(I_{\text{prev}}, I_{\text{curr}})$ \label{alg:cond-superset} \\
        \quad or $I_{\text{prev}}$ equals $I_{\text{curr}}$ \label{alg:cond-equal}
      }
        \State $G \gets G \cup \{E_i\}$ \label{alg:add-ei}
      \Else
        \State append $G$ to $\mathcal{F}$ \label{alg:append-g}
        \State \Return $\mathcal{F} \cup \textsc{GreedyFusion}(C[i:], N - i)$ \label{alg:return-recurse}
      \EndIf

      \State $I_{\text{prev}} \gets I_{\text{curr}}$ \label{alg:update-iprev}
      \State $R_{\text{up}} \gets R_{\text{dwn}}$ \label{alg:update-rup-loop}
    \EndFor \label{alg:loop-end}

    \State \Return $\mathcal{F} \cup \{G\}$ \label{alg:return-final}

  \end{algorithmic}
\end{algorithm}

\begin{figure}
  \centering

  \begin{subfigure}[t]{.4\columnwidth}
    \centering
    \caption{Iteration spaces}
    \label{fig:greedy-iter-spaces}
    \begin{align*}
      \fcolorbox{blue}{fusionblue}{$E_1 : [M,N,K]$} \\
      \fcolorbox{blue}{fusionblue}{$E_2 : [M,N,P]$} \\
      \fcolorbox{blue}{fusionblue}{$E_3 : [M,N,P,Q]$} \\
      \fcolorbox{yellow}{fusionyellow}{$E_4 : [M,N,Q]$} \\
      \fcolorbox{yellow}{fusionyellow}{$E_5 : [N]$}
    \end{align*}
  \end{subfigure}
  \hfill
  \begin{subfigure}[t]{.4\columnwidth}
    \centering
    \caption{Pairwise intersections}
    \label{fig:greedy-intersections}
    \begin{align*}
      \fcolorbox{blue}{fusionblue}{$E_1 \rightarrow E_2 : [M,N]$} \\
      \fcolorbox{blue}{fusionblue}{$E_2 \rightarrow E_3 : [M,N,P]$} \\
      \fcolorbox{yellow}{fusionyellow}{$E_3 \rightarrow E_4 : [M,N,Q]$} \\
      \fcolorbox{yellow}{fusionyellow}{$E_4 \rightarrow E_5 : [N]$}
    \end{align*}
  \end{subfigure}

  \begin{subfigure}[t]{\columnwidth}
    \centering
    \caption{Final mapping into two fusion groups}
    \label{fig:greedy-fusion-groups}

    \begin{mdframed}[roundcorner=4pt]
      \lstset{style=compactPython}
      \begin{lstlisting}
(*@\fgone{\ttfamily\# Fusion Group 1 (E1--E3)}@*)
for n in range(N):
  for m in range(M):
    for k in range(K):               # E1
      Z[m,n] += A[m,k] * B[k,n]
    for p in range(P):               # E2
      Y[m,n,p] = Z[m,n] * C[p]
      for q in range(Q):             # E3
        X[m,n,q] += Y[m,n,p] * W[q]
(*@\fgtwo{\ttfamily\# Fusion Group 2 (E4--E5)}@*)
    # Q-fiber of X is ready here (transition to E4)
    for q in range(Q):               # E4
      V[n] += X[m,n,q] * D[q]
  U[n] = f(V[n])                     # E5
      \end{lstlisting}
    \end{mdframed}
  \end{subfigure}

  \caption{
    Example greedy fusion over a five-Einsum cascade. Blue: Fusion Group 1, Yellow: Fusion Group 2.
    (a) Iteration spaces of Einsums 1--5.
    (b) Corresponding pairwise intersections.
    (c) Resulting fused mapping into two fusion groups.
  }
  \label{fig:greedy-fusion}
\end{figure}

We present our fusion strategy in Listing~\ref{alg:greedy-fusion}.
For simplicity, we assume the cascade is a sequential DAG of Einsums.
Given two Einsums, fusion is always possible according to Section~\ref{sec:pairwise-fusion}.
Additionally, the pairwise classification in Section~\ref{sec:pairwise-fusion} informs which traversal orders are allowed when greedily stitching.

Given a cascade, the greedy algorithm fuses the first two Einsums, forming a fusion group.
It then intersects their iteration spaces to determine common ranks, if any.
To determine if the next Einsum can join this fusion group, line~\ref{alg:get-rdwn-i} retrieves its iteration space.
Line~\ref{alg:get-icurr} then intersects this iteration space with the last Einsum in the current fusion group.
If this intersected set is equal to, a subset of, or a superset of the intersected set of the first two Einsums (lines~\ref{alg:cond-equal}, \ref{alg:cond-subset}, and~\ref{alg:cond-superset}), the algorithm adds the new Einsum to the fusion group.
The process repeats with the next Einsum in the cascade.
Otherwise, the current fusion group ends and the process repeats starting from the new Einsum.
The pairwise intersection lists are the byproduct of output-input stationarity from pairwise fusion: ranks that survive intersection \emph{must} appear at stationary levels of the final traversal order.
Figure~\ref{fig:greedy-fusion} depicts an example cascade and the resulting fusion groups after greedy stitching. Since $N$ is shared across all Einsums, the mapping must be $N$-stationary to enable output-upstream/input-downstream stationarity.

Greedy stitching is one approach.
Another approach is to globally form the pairwise intersected lists for every pair of (dependent) Einsums in a cascade.
The stitching algorithm can then select the group of Einsums that form the longest ``passing'' set of pairwise intersections.

%% file: sections/5-mamba_fusion.tex
\section{Fusing Mamba}\label{sec:mamba_fuse}
\begin{figure*}
  \centering
  \includegraphics[width=0.9\textwidth]{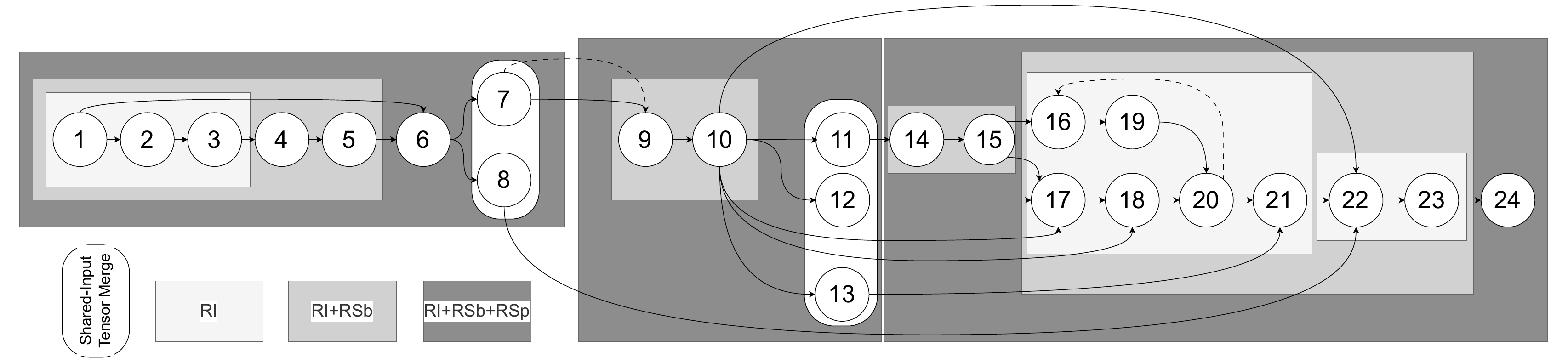}
  \caption{Fusion opportunities in the Mamba cascade.
  Each node denotes an Einsum, and directed edges indicate data dependencies
  between their input and output tensors. Dashed arrows indicate recurrent or correlation Einsums.}
  \label{fig:einsum-dependencies}
\end{figure*}

\begin{figure}
\begin{centering}
    \includegraphics[width=\columnwidth]{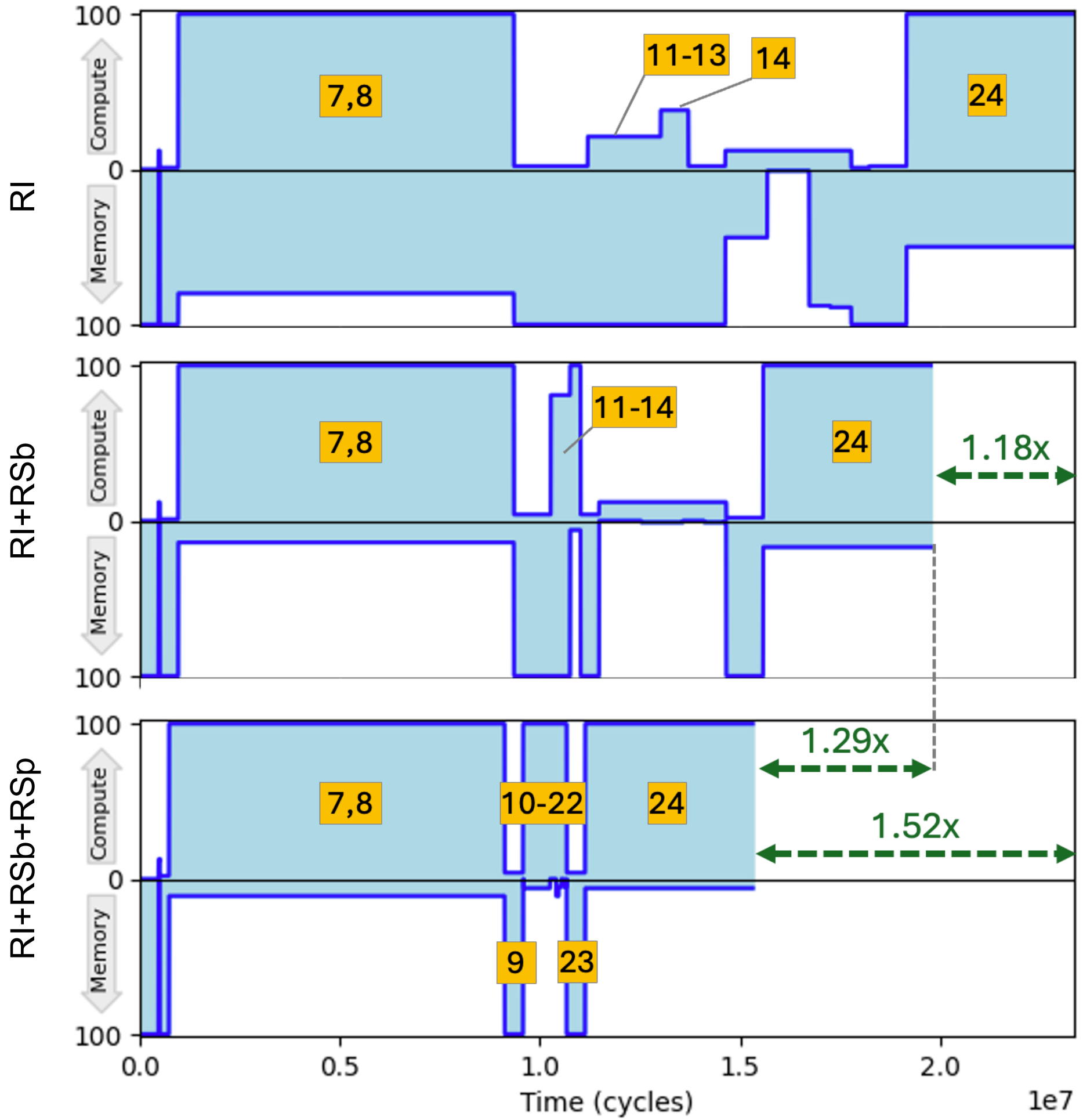}
    \caption{Roofline utilization over time of single Mamba layer for different fusion strategies. All assume algorithmic minimum accesses on intra-Einsum traffic. Successively applying fusion variants reduces the overall number of fusion groups and improves overall latency.}

    \label{fig:mamba:rank-category}
\end{centering}
\end{figure}

In Mamba, fusion opportunities exist everywhere there is an outgoing arrow to a box or operand in Figure~\ref{fig:cascade}.
We now leverage the Einsum fusion taxonomy to apply stitching in a systematic manner.

Before stitching, we first apply an algorithmic transformation to the cascade: shared-input tensor merging.

This is a common optimization strategy often used to pack multiple GEMM operations into a single, larger GEMM computation (~\cite{Gu:2023:MLT,Zheng:2023:BLD,Ansel:2024:Pytorch}).
We apply shared-input merging on (A)~$\NEX$ to produce both $\TX$ and $\RX$ simultaneously (Einsums 7--8, Figure~\ref{fig:cascade}), (B)~$X$ to produce $B$, $C$, and $TT\Delta$ (Einsums 12--14), and (C)~$\Delta$ to produce $\bar{A}$ and $\bar{B}$ (Einsums 16--17).

Given this algebraic transformation, we fuse Mamba by stitching in four different ways:

\subsection{RI-Only Stitching}
Rank-isomorphic fusion maintains the same iteration space between a pair of Einsums.
In this mode, we modify our greedy stitching algorithm (Algorithm~\ref{alg:greedy-fusion}) to fuse a downstream Einsum into a fusion group \emph{when the iteration space of that Einsum is isomorphic to the upstream} (apply the condition on line~\ref{alg:cond-equal} only).

Figure~\ref{fig:einsum-dependencies} collapses the Einsums (see Figure~\ref{fig:cascade}) into nodes.
Light grey boxes show the fusion groups formed by RI-only stitching.
In total, we reduce the number of fusion groups from 24 (unfused, individual Einsums) to 12.

Figure~\ref{fig:mamba:rank-category} (top) shows the utilization diagram of ideal rank isomorphic (RI) fusion on Mamba.
In particular, RI fusion significantly reduces the latency of the SSM portion of Mamba (Einsums 16--21) compared to a best-case, unfused mapping.

\subsection{RI+RSb Stitching}
The transition from producing NUM (Einsum 3), which reduces over the $\ED$ rank, to producing $\SQEX$\@(Einsum 5) can leverage RSb fusion.
We expand the RI stitching approach to now include the cases when the pairwise intersection with the downstream Einsum is a subset of the previous pairwise intersection (Lines~\ref{alg:cond-equal}--~\ref{alg:cond-subset}).

Figure~\ref{fig:einsum-dependencies} shows the fusion groups formed by RI+RSb stitching.
The total number of fusion groups is now eight.
In particular, note that the SSM region (Einsums 16-21) can now directly pass its output ($S$, Einsum 21) with the post-processing that produces $Y$ (Einsums 22--23).

Figure~\ref{fig:mamba:rank-category} (middle) shows the utilization diagram of ideal RI+RSb fusion on Mamba. Adding RSb has the potential to improve performance by 1.18x compared to RI-only.

\subsection{RI+RSb+RSp Stitching}
Finally, fully applying the greedy stitching algorithm, (Algorithm~\ref{alg:greedy-fusion}) results in RI+RSb+RSp stitching.

In Mamba, an RSp opportunity exists between the $\NEX$ Einsum and the $\TX$ Einsum (Einsums 5--6).
Figure~\ref{fig:mamba:rank-category} (bottom) shows the utilization diagram of ideal rank-supersetted fusion with RSb and RI fusion.
Adding RSp reduces the number of fusion groups to three.
In particular, Einsums 11-13, a shared-input tensor GEMM with non-ideal aspect ratios, now achieves 100\% compute throughput utilization as its inputs now remain on-chip.
Likewise, some portions of the SSM---traditionally considered memory-bound---now achieve maximum compute utilization as well (Einsums 16--22).

\subsection{Fully Fused (RI+RSb+RSp+RD Stitching)}
The greedy stitching algorithm (Algorithm~\ref{alg:greedy-fusion}) cannot successfully apply RD fusion to Mamba, as doing so breaks the output-upstream/input-downstream stationary requirements defined earlier.
Prior work simply recomputes the data as needed when RD fusion is unavailable for stitching~\cite{gilbert:2024:LEF}, or \emph{weakly} fuses the data such that intermediates are not guaranteed to remain on-chip. We take a different approach.

In Mamba, an opportunity for RD fusion exists between the first and second fusion groups of RI+RSb+RSp fusion (previous section) as well as between the second and third fusion groups.
Rather than recomputing, we carefully form tiles of intermediate data, allow partial products of the upstream intermediate tensor to write to main memory, then trigger the downstream Einsum on the final write of an element (or tile) in the upstream.
In Figure~\ref{fig:greedy-fusion-groups}, note that $X$ cannot be fully fused with the $V$ Einsum as an entire $Q$ fiber is produced at a time. Rather than waiting for the entire fiber, our strategy begins executing $V$ as soon as a \emph{final} element of $X$ is ready.

We term this fusion strategy ``fully fused'' as it creates one fusion group.
This requires careful binding to the underlying architecture (see Section~\ref{sec:eval}).

\subsection{Fusing with Iterative rank $I$}

We find partitioning aids in fusing Einsums with generational ranks.
If an Einsum contains generational ranks (such as $\TTX$ and the SSM), we can unroll the cascade and apply the above strategies to the unrolled cascade.
However, this unrolling implies a loop order that quickly varies along the iterative rank in order to keep a reasonable-sized portion on-chip.
For example, if we are $I$-stationary on $H$, we must store a $B \times D \times N$ partition on-chip.
However, if we are $B$--$D$--$N$-stationary, only a unit-sized element of $H$ stays on-chip with a guarantee that there will be no spills to main memory.
Partitioning along the iterative rank ($I$) can aid in keeping larger tiles of the iterative rank on-chip, reducing the requirement for a $B$--$D$--$N$-stationary dataflow.
This is the strategy we employ for Mambalaya's fully fused implementation (Sections~\ref{sec:eval},~\ref{sec:methodology}).

%% file: sections/7-binding.tex
\section{Mambalaya Accelerator}\label{sec:eval}
To both better support low-intensity operations and enable fusion with high-intensity operations, we propose the following \emph{reconfigurable} architecture for Mambalaya.
Overall, \emph{Mambalaya's flexible architecture enables the proposed fusion mappings}.
Without a flexible, reconfigurable 2D-array containing both low- and high-intensity compute units, RI, RI+RSb, RI+RSb+RSp, and the fully fused mappings would not be able to execute efficiently.

\subsection{Mambalaya Architectural Overview}
\begin{figure}
  \centering
  \begin{subfigure}[t]{0.70\columnwidth}
    \centering
    \includegraphics[width=\linewidth]{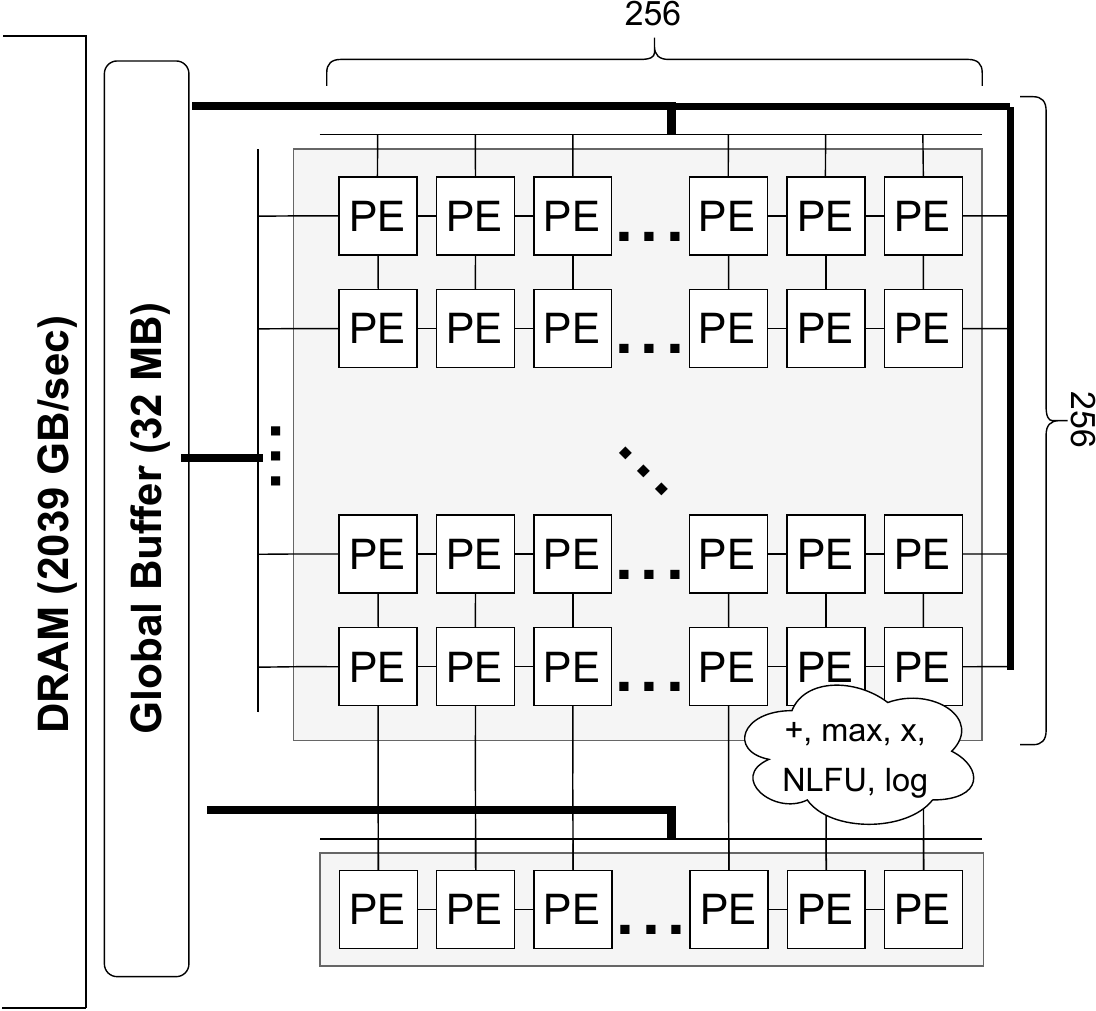}
    \caption{2D mode}
    \label{fig:mamba:arch}
  \end{subfigure}\\
  \begin{subfigure}[t]{0.7\columnwidth}
    \centering
    \includegraphics[width=\linewidth]{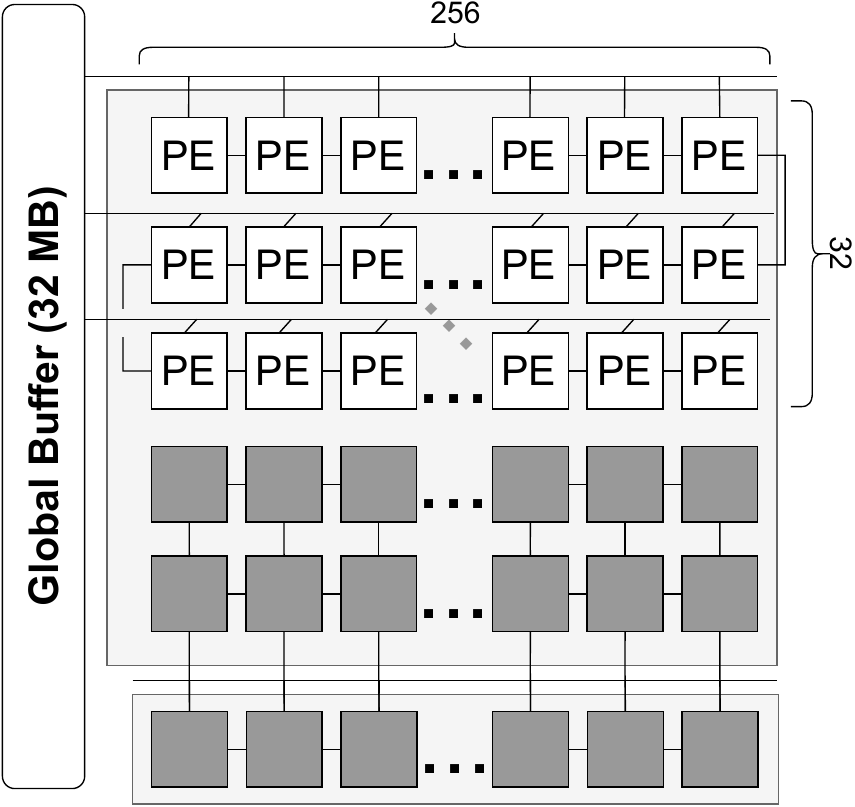}
    \caption{1D mode}
    \label{fig:mamba:arch-1D}
  \end{subfigure}
  \caption{Mambalaya’s reconfigurable architecture, with the (a) 2D array in 2D mode and (b) the 2D array in 1D mode. The 1D 256-PE array also connects to the first row of the 2D array.
  }
  \label{fig:mamba:arch-both}
\end{figure}

 Mambalaya consists of a \emph{2D PE array} with a \emph{reconfigurable} network that has two modes: (1)~A 2D mode where all PEs are connected in a 2D structure with the store-and-forward network assumed by both TPU~\cite{tpu-v2-v3} and FuseMax~\cite{Nayak:2024:FML_micro}; and a (2)~1D mode where a \emph{subset} of the PEs (8192 PEs) are directly connected to the global buffer and linearly to each other.
 Figure~\ref{fig:mamba:arch} shows the 2D mode, while Figure~\ref{fig:mamba:arch-1D} shows the 1D mode.
 Mambalaya enters the 2D mode when either (1)~executing a GEMM Einsum or (2)~executing a fusion group that contains both GEMM and low-intensity Einsums.
 When executing a fusion group with only low-intensity Einsums, Mambalaya enters 1D mode.
 The 1D mode with 8192 PEs is necessary to ensure the number of available compute units does not limit low-intensity computation.

To enable elementwise fusion, all PEs contain both high-intensity (MACCs) and low-intensity (log/max/SiLU/exp) functional units.
Each PE contains a 6-stage, \emph{pipelined} functional unit, allowing an operation to complete in each cycle.
We leverage MARCA's non-linear function unit for SiLU and the exponential function~\cite{Li:MARCA}, and the logarithm implementation from Mal\'{\i}k~\cite{malik:2017:dhc}.

We additionally include a low-intensity 1D PE array (256 PEs) directly connected to the global buffer and to the first and last rows of the 2D structure.
We use this structure when a given mapping fuses a low-intensity, producer Einsum with a high-intensity, consumer Einsum: the 1D array broadcasts its result to the 2D array.

\subsection{Mapping and Binding to Mambalaya}

\par{\textbf{RI-only}}: We bind all fusion groups containing elementwise operations to the 2D PE array in 1D mode (8192 PEs).
We bind GEMM computations to the 2D array (in 2D mode).

\par{\textbf{RI+RSb}}: In this variant, fusion groups consist of either elementwise or linear operations, followed by an elementwise operation.
In the latter case, we run GEMMs followed by elementwise operations on the \emph{2D} array in its 2D mode (Einsums 14--15 in Figure~\ref{fig:einsum-dependencies}).
Since the GEMM result is already on the 2D array, the subsequent elementwise operation can remain in 2D mode.

\par{\textbf{RI+RSb+RSp} and \textbf{Fully Fused}}:\label{par:1d-broadcast} Fusion groups now consist of (1)~elementwise operations followed by a GEMM, or (2)~a GEMM or linear operation followed by elementwise operations.
In the first case, the elementwise Einsums cannot run in the 1D mode of the 2D array, as the GEMM computation needs all 256$\times$256 elements of the 2D array. Thus, all elementwise operations preceding GEMMs in a fusion group are bound to the smaller, 1D-only array (Einsums 1--6 in Figure~\ref{fig:einsum-dependencies}).
The resulting intermediate is then broadcast to the 2D PE array for use in the GEMM operations.
Any elementwise Einsum that follows a GEMM will execute in 2D mode.

%% file: sections/8-evaluation.tex
\section{Methodology and Evaluation}\label{sec:methodology}
\begin{figure*}
    \centering
    \includegraphics[scale=1, width=\textwidth]{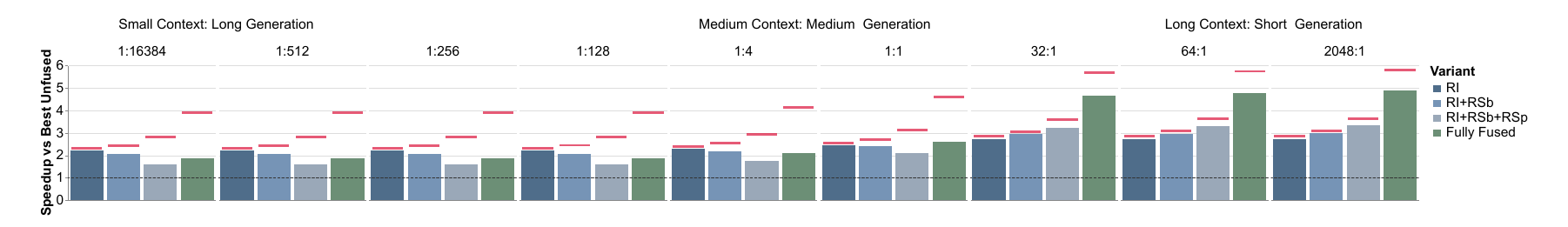}
    \caption{End-to-end performance across the different variants for the \texttt{mamba-370m} model~\cite{hf:2024:mamba2.8b}. In red is the ideal performance (best-case scenarios with algorithmic minimum accesses). Bars indicate the \emph{achieved} performance.
    }
    \label{fig:end-to-end}
\end{figure*}

\begin{figure*}
    \centering
    \includegraphics[width=\textwidth]{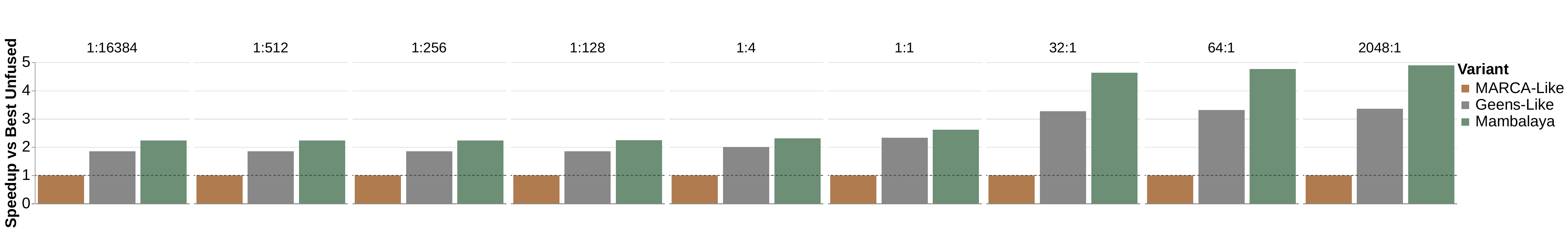}
    \caption{Best Mambalaya variant compared to two state-of-the-art Mamba accelerators: MARCA~\cite{Li:MARCA} and Geens et al.~\cite{geens2025finegrainedfusion}. Mambalaya achieves 4.9$\times$ and 1.5$\times$ speedup against MARCA-like and Geens-Like, respectively.}
    \label{fig:sota}
\end{figure*}

\subsection{Evaluation Setup}
\textbf{Accelerator Modeling.}
To model Mambalaya, we leverage Timeloop, a tensor algebra accelerator modeling and design exploration tool~\cite{timeloop}.
We use Timeloop to simulate the performance of individual Einsums in the cascade.

For each fusion strategy, we specify the mapping constraints imposed by Algorithm~\ref{alg:greedy-fusion} and feed said constraints into the Timeloop mapper for each individual Einsum.
The mapper searches the mapping space and returns a pseudo-optimal mapping along with the corresponding memory and compute costs.

\textbf{Workload Parameters.}
Given the Mamba workload, the only user-specified ranks are the batch size $B$, the prefill length ($I > 1$), and the token generation length ($I=1$).
Following FLAT~\cite{flat} and FuseMax~\cite{Nayak:2024:FML_micro}, we use a batch size of 64. We run on both models \verb|mamba-370m| and \verb|mamba-2.8b|~\cite{hf:2024:mamba2.8b}. The larger model more than doubles the $E$ and $D$ ranks ( Figure~\ref{fig:cascade}) and uses 64 layers instead of 48.
We vary the token length from $I=1$ (token generation) to $I=2^{20}$, in keeping with common configurations in prior work~\cite{flat, Nayak:2024:FML_micro}.

\textbf{Hardware Configurations.}
Where possible, we design Mambalaya to match the configurations of a single NVIDIA H100 GPU\@; otherwise we choose parameters with lower values than the GPU\@.
In this way, we ensure we are at most iso-area with the H100.
Table~\ref{tab:hw_config} summarizes the configurations.

\begin{table}
  \caption{Mambalaya configuration compared to  H100~\cite{nvidia:2022:GWH}}
  \label{tab:hw_config}
  \centering
  \footnotesize
  \begin{tabular}{@{}llcc@{}}
    \toprule
    \textbf{Category} & \textbf{Feature} & \textbf{H100 GPU} & \textbf{Mambalaya} \\
    \midrule
    Compute Units
        & \# FP16 CUDA Cores      & 14{,}592       & ---       \\
        & \# Tensor Cores         & 456            & ---       \\
        & Total \# PEs            & ---            & 65{,}536 + 256\\
        & 1D PE Config. (2D)    & ---            & 8192$\times$1 \\
        & 2D PE Config. (2D)  & ---            & 256$\times$256 \\
    \midrule
    Frequency
        & Clock Frequency         & 1.75\,GHz      & 1.75\,GHz  \\
    \midrule
    Memory
        & Memory Bandwidth        & 2039\,GB/s     & 2039\,GB/s \\
        & L2 / Global Buffer      & 50\,MB         & 32\,MB     \\
        & Total Register Size     & 28.5\,MB       & 4.25\,MB   \\
    \bottomrule
  \end{tabular}
\end{table}

\subsection{Prior State-of-the-Art}
MARCA~\cite{Li:MARCA} is the prior state-of-the-art accelerator for Mamba.
Although recently released, the extended Einsum framework, combined with our proposed fusion taxonomy, enables us to quickly and easily characterize their work.

Within the SSM region of the Mamba cascade, MARCA applies RI fusion to the back-to-back elementwise Einsums.
However, MARCA uses non-unit size intermediate tensors, making the implementation brittle to changes in on-chip buffer sizes (see Table~\ref{tab:relatedworks}).
To address this, Geens et al.,~\cite{geens2025finegrainedfusion} propose a fine-grained, memory-aware fusion strategy that partitions the $H$ state tensor to unit size along the $I$ rank.
When on-chip memory is small, they further tile along the $D$ and $N$ ranks of $H$ to enable the state tensor to fit on-chip.

To enable an apples-to-apples comparison, we give both baselines the benefit of the doubt: we assume best-case unfused Einsums with algorithmic minimum traffic, and apply rank-isomorphic fusion to the SSM region (Einsums 16-21 for Mamba-1) on the Mambalaya architecture. This isolates fusion and mapping strategy as the independent variable.
We call these design points \emph{MARCA-like} and \emph{Geens-like}.

\subsection{Evaluating Mambalaya}

\begin{figure}
    \centering

    \begin{subfigure}[t]{.5\textwidth}
        \centering
        \includegraphics[
            width=\linewidth
        ]{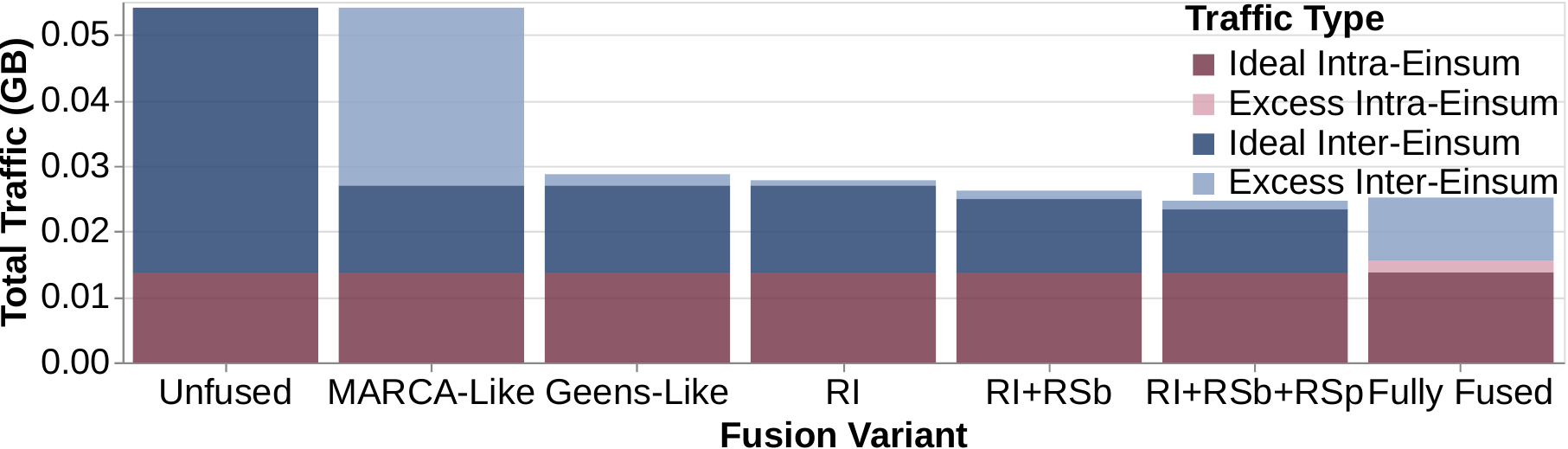}
        \caption{Prefill phase (B=64, I=2048).}
        \label{fig:traffic:prefill}
    \end{subfigure}

    \begin{subfigure}[t]{.5\textwidth}
        \centering
        \includegraphics[
            width=\linewidth
        ]{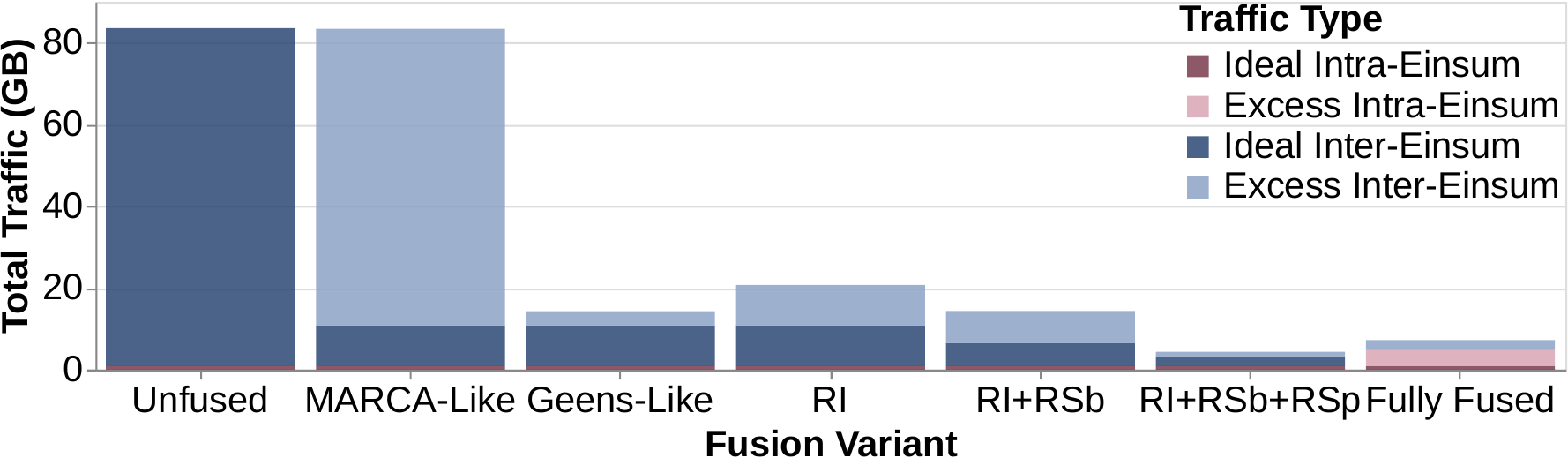}
        \caption{Decode / Token generation (B=64, I=1).}
        \label{fig:traffic:decode}
    \end{subfigure}

    \caption{
        Inter- and intra-Einsum traffic across each fusion variant.
        Dark red and dark blue indicate the \emph{ideal} traffic for that variant,
        while lighter colors show the excess achieved in practice.
        The best-case RI variant is used as the ideal baseline for both
        MARCA-Like~\cite{Li:MARCA} and
        Geens-Like~\cite{geens2025finegrainedfusion}
    }
    \label{fig:traffic}
\end{figure}

\begin{figure}
    \centering

    \begin{subfigure}[t]{0.49\columnwidth}
        \centering
        \includegraphics[width=\linewidth]{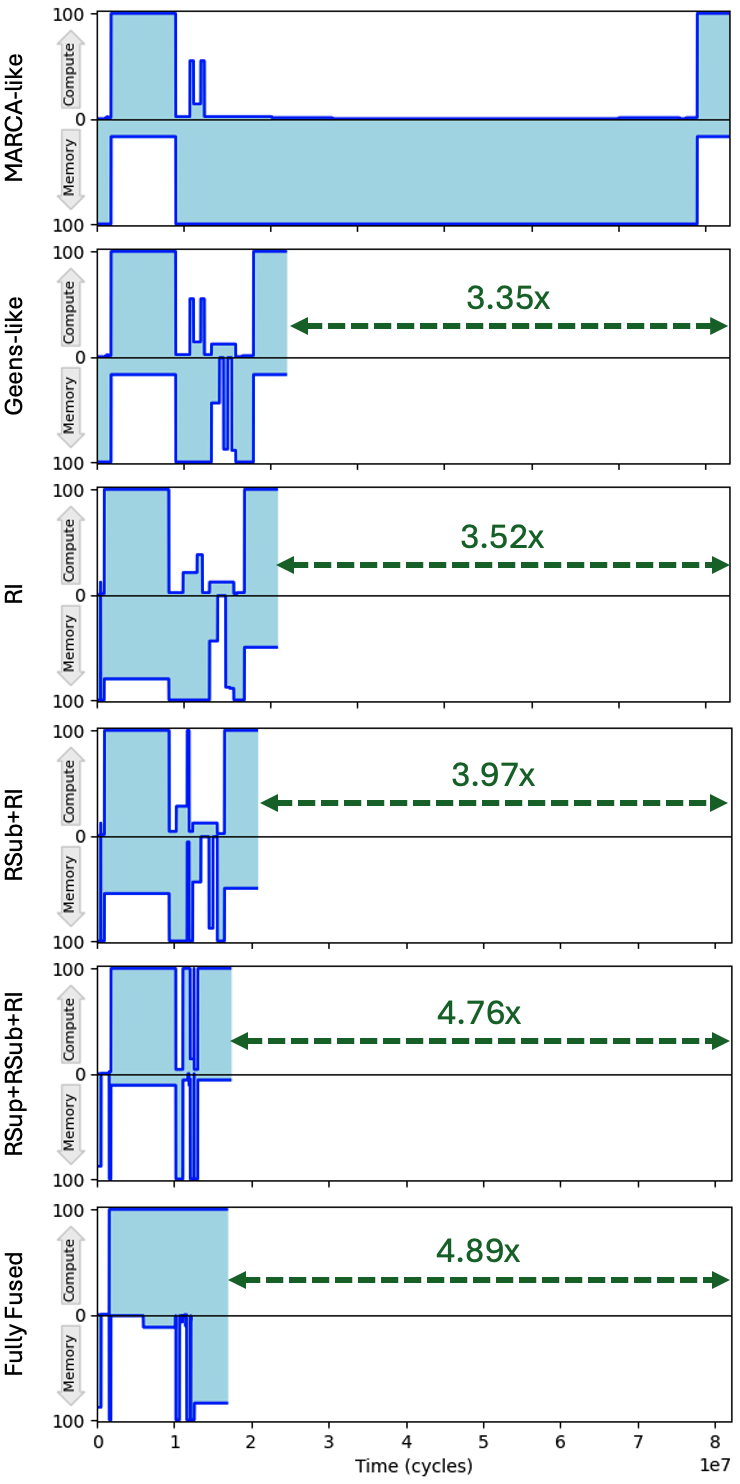}
        \caption{Prefill.}
        \label{fig:zoom_in_prefill}
    \end{subfigure}
    \begin{subfigure}[t]{0.49\columnwidth}
        \centering
        \includegraphics[width=\linewidth]{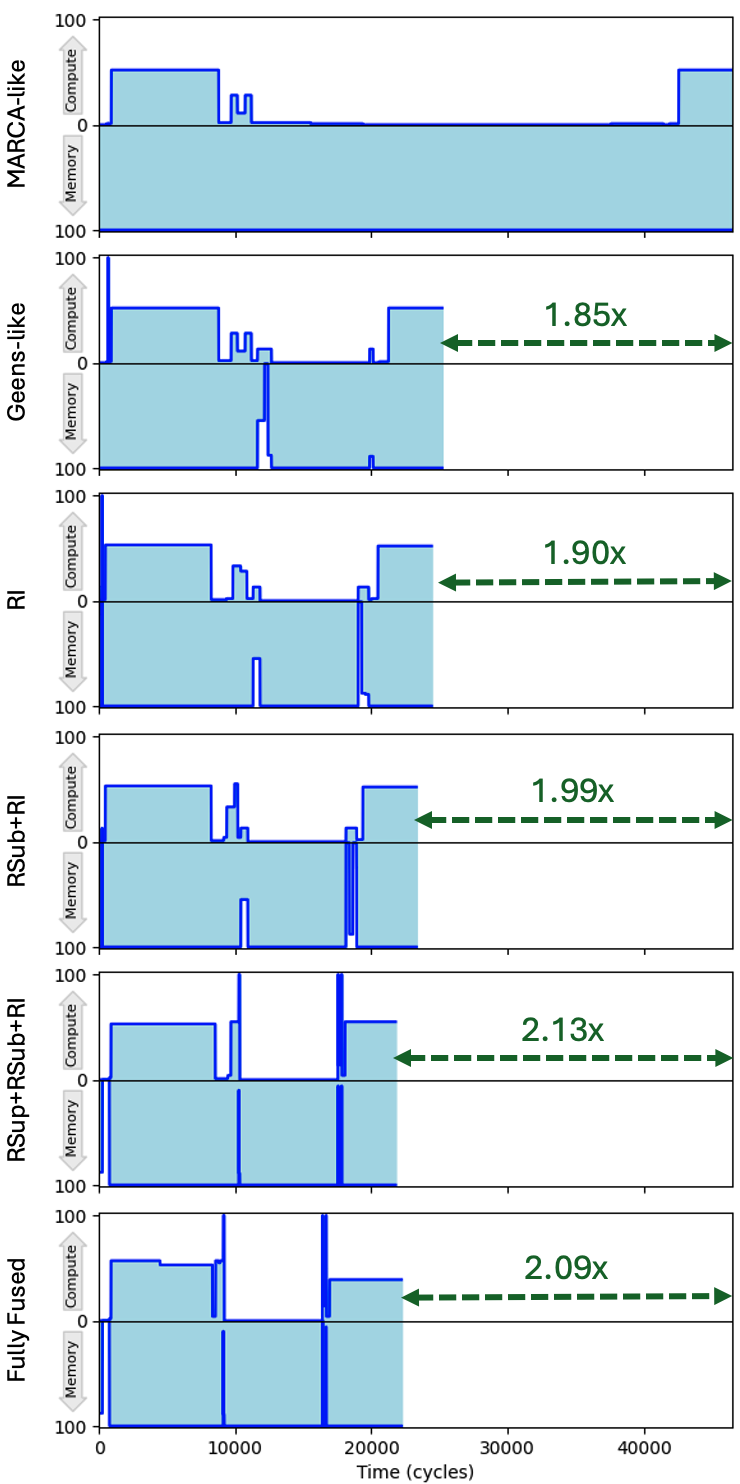}
        \caption{Token generation.}
        \label{fig:zoom_in_decode}
    \end{subfigure}

    \caption{Roofline utilization over time for both prefill and token generation phases.}
    \label{fig:zoom_in_combined}
\end{figure}

\subsubsection{End-to-End Performance of Fusion Variants}
Figure~\ref{fig:end-to-end} shows the end-to-end performance of Mambalaya.
Each bar grouping is a specific ratio of context length to token generation length.
We show three common scenarios: a small context but long token generation (e.g., requesting an explanation on a topic), a medium-length context to similarly-sized token generation (e.g., asking the LLM to edit a paragraph), and a large context to short token generation (e.g., summarizing a document).

Regardless of prefill and decode length, performance is consistent across similar \emph{ratios} (of prefill to  decode).
For relatively large decode length, RI fusion performs the best, achieving close to its ideal at 2.23$\times$ over the unfused baseline.
During token generation---which dominates small context to large generation scenarios---only the rank-isomorphic strategy is able to bind to the 1D configuration (8192 PEs) of the 2D array during the normalization steps (Einsums 1--6).
Meanwhile, all other strategies are resource limited as they use the 1D array (256 PEs) in order to later broadcast their results to the 2D Einsums.
Thus, RI fusion performs well during token generation.

As the sequence length in prefill increases relative to the decode length, the fully fused approach dominates performance with a speedup of $4.9\times$ over the unfused baseline. RI, RI+RSb, and RI+RSb+RSp have speedups of 2.72$\times$, 2.99$\times$, and 3.35$\times$, respectively.
With parallel pipelining, performance improves to $3.9\times, 4.7\times, 5.9\times, 6\times$ during prefill for RI, RI+RSb, RI+RSb+RSp, and fully fused, respectively.

All three strategies are near their ideal limit (red line).
The fully fused approach does not reach its ideal limit as tensors $X$ and $\LEX$ (Einsums 1 and 10) need two passes (see pass analysis in FuseMax~\cite{Nayak:2024:FML_micro}) and thus, must be loaded multiple times.
Reducing the number of passes on these tensors is not possible as they are used in both reduction Einsums (reducing over the $D$ rank, Einsums 2 and 12--13), and broadcast/elementwise Einsums (Einsums 6, 17, 22).
Additionally, we send $\RX$ (Einsum 8) off-chip, as it has a long dependency chain: it is not needed again until Einsum 22.
This frees up buffer space for other tensors during execution.

\subsubsection{Performance Compared to the Prior State-of-the-Art}
Figure~\ref{fig:sota} compares the best Mambalaya implementation against the MARCA-Like and Geens-Like accelerators.
Since all accelerators minimize the memory accesses (and corresponding bottle-necked latency) of the $H$ tensor, they all see significant improvement over the unfused baseline.
In particular, in large context, shorter token generation scenarios (e.g., summarizing a document), Mambalaya achieves a 4.9$\times$ speedup over the unfused baseline, and provides greater than 44$\%$ improvement over the state-of-the-art accelerators.

\subsubsection{Intra-/Inter-Einsum Traffic}
Although we assumed so in the algorithmic minimum access case in Section~\ref{sec:fusion}, intra-Einsum traffic is not free.
Figure~\ref{fig:traffic} shows the breakdown of intra-Einsum traffic to inter-Einsum traffic across all variants (for a single layer) in both prefill and decode.
As previously noted, an unfused implementation has excessive inter-Einsum traffic, as it accesses its input and output tensors from main memory.
Fusion drastically reduces this: all variants successfully reduce the inter-Einsum traffic ranging from 4$\times$ to 34$\times$ improvement. All variants---except fully fused---achieve near-perfect intra-Einsum traffic, as the tensors that must be loaded on-chip regardless of fusion (i.e., weight tensors) are relatively small.
The fully fused variant severely constrains the intra-Einsum dataflow of all Einsums in the cascade, producing large intermediate tensor partial products. This leads to comparatively worse intra-Einsum traffic (light pink segment).

\subsubsection{Utilization Analysis}
Figure~\ref{fig:zoom_in_combined} shows the roofline utilization over time for each state-of-the-art baseline and our four proposed fusion strategies, respectively.
In prefill, moving from RI to the fully fused mapping, each fusion strategy successfully reduces overall memory volume (shaded area of memory bandwidth region).
Geens-like only fuses in the SSM region, yet this is enough to provide a 3.35$\times$ improvement over the MARCA-like baseline.
From Geens-like to RI-only fusion, the number of fusion groups reduces by 2. This amount does not provide a significant benefit for RI fusion over Geens-like.
When we start fusing broadcast operations as well (RI+RSb), a slight improvement in compute utilization for the fused regions occurs---specifically, the GEMM followed by an elementwise Einsum (Einsums 14--15) now achieves full compute utilization.

When stitching includes RSp, nearly all memory-bound regions in the previous strategies disappear, with a 4.76$\times$ improvement over the MARCA-like baseline.
The fully fused variant, although it consists of one large fusion group, performs marginally better than RI+RSb+RSp.
Notice the increase in memory volume in Figure~\ref{fig:zoom_in_prefill} compared to the RI+RSb+RSp variant, particularly in the last phase where computation is a GEMM.
The increased inter-Einsum traffic in Figure~\ref{fig:traffic:prefill}, points to the increase in partial products as the underlying cause.
During prefill, the benefit of achieving a smooth, compute-bound execution in full fusion outweighs the increase in memory traffic.
However, during token generation (Figure~\ref{fig:zoom_in_decode}), the problem size is not large enough to fully utilize the compute units.
Thus, the increase in memory traffic now becomes a bottleneck for the fully fused variant, resulting in relatively poor latency.

Overall, Mambalaya successfully achieves a better compute utilization compared to the state-of-the-art (Figure~\ref{fig:zoom_in_combined}). Additionally, Mambalaya reaches a per-layer performance improvement of 4.9$\times$ and 1.5$\times$ over MARCA-like and Geens-like. In end-to-end scenarios with varying context-length to token generation, Mambalaya achieves a geomean speedup of 3$\times$ and 1.3$\times$ over MARCA-like and Geens like respectively.

%% file: sections/10-conclusion.tex
\section{Conclusion}
Mamba is a complex application and requires the development of a complex set of optimization strategies.
In this paper, we use this complexity as motivation to advance the state-of-the-art in fusion techniques for tensor algebra applications that can be expressed as Einsums. We then leverage these to design Mambalaya, a reconfigurable, flexible architecture that is the first custom accelerator to support the execution of a fully-fused Mamba across the entire cascade.

Although these techniques have been motivated by Mamba, they are quite general. By taxonomizing the space of fusion, we can now mechanically apply a systematic set of fusion strategies to any tensor workload, no matter the complexity. Our proposed strategies---rank-isomorphic, rank-subsetted, rank-supersetted, and rank-disjoint---can be combined to fuse any Einsum cascade, simply by observing their tensor indices.

All together, these techniques can reduce intermediate tensor traffic to near-minimum. This enables a robust and promising area of future work that carefully explores and trades off inter- and intra- tensor reuse via automatic design-space exploration. By leveraging Einsums, we make this work naturally available for inclusion by a large number of existing tools for design space exploration within the computer architecture community.

%% file: sections/appendix.tex
\section{Mamba-1 Cascade}

\definecolor{myblue}{RGB}{0,102,204}
\definecolor{mygreen}{RGB}{0,153,0}

\subsection{Rank Names}

\subsubsection*{Word/token length ranks}
\begin{itemize}
  \item $I$ --- sequence length
  \begin{itemize}
    \item $I_0$ --- sequence length of the prompt
  \end{itemize}
\end{itemize}

\subsubsection*{Vocabulary/embedding ranks}
\begin{itemize}[leftmargin=1.5em]
  \item $V$ --- dictionary size (words in vocabulary)
  \item $\ED$ --- input global space embedding
  \item $D$ --- channel, expanded from $\ED$
  \item $R$ --- embedding space for the time step tensor ($\Delta$)
\end{itemize}

\subsubsection*{Other Ranks}
\begin{itemize}[leftmargin=1.5em]
  \item $N$ --- hidden state/memory dimension. The larger this is, the better the hidden state tensor is able to remember context
  \item $B_1$ --- number of batches
  \item $B_0$ --- size of a batch
  \item $B$ --- total number of prompts \quad ($B = B_1 \times B_0$)
  \item $L$ --- number of layers
\end{itemize}

\subsection*{Input}

\subsubsection*{Input tensors}
\begin{itemize}[leftmargin=1.5em]
  \item \textcolor{myblue}{\(\mathrm{Tokens}^{I,V}\)} --- raw input (input to first layer only)
  \item \(\LEX^{I,ED}\) --- input after embedding (input to subsequent layers)
\end{itemize}

\subsection*{Principal Intermediate Tensors}
\begin{itemize}[leftmargin=1.5em]
  \item \(\Delta^{I,D}\) --- time step for discretization
  \item \(\NEX^{I,ED}\) --- normalized input
  \item \(X^{I,D}\) --- input to state-space model
  \item \(\bar{A}^{I,D,N}\) --- discretized \(A\) tensor
  \item \(\bar{B}^{I,D,N}\) --- discretized \(B\) tensor
  \item \(H^{I,D,N}\) --- hidden state/memory tensor
\end{itemize}

\subsection*{Weight and Trained Tensors}
\begin{itemize}[leftmargin=1.5em]
  \item \textcolor{mygreen}{\(\WEX^{ED}\)} --- weight tensor to create \(\NEX\)
  \item \textcolor{mygreen}{\(\WTTX^{ED,D}\)} --- weight tensor to create \(\TTX\)
  \item \textcolor{mygreen}{\(\WRX^{ED,D}\)} --- weight tensor to create \(\RX\)
  \item \textcolor{mygreen}{\(\WTX^{D,F}\)} --- weight tensor (convolution filter) to create \(\TX\)
  \item \textcolor{mygreen}{\(\WB^{D,N}\)} --- weight tensor to create \(B\) (continuous)
  \item \textcolor{mygreen}{\(\WC^{D,N}\)} --- weight tensor to create \(C\) (continuous)
  \item \textcolor{mygreen}{\(\WTT\Delta^{D,R}\)} --- weight tensor to create \(\TT\Delta\) (for \(\Delta\))
  \item \textcolor{mygreen}{\(\WT\Delta^{R,D}\)} --- weight tensor to create \(T\Delta\) (for \(\Delta\))
  \item \textcolor{mygreen}{\(\WEY^{D,ED}\)} --- weight tensor to project back to the embedding space
  \item \textcolor{mygreen}{\(\WET^{D,ED}\)} --- weight tensor to project from the embedding space to logits (also projects from token space to embedding space)
\end{itemize}

\noindent\textit{Note:} Superscripts denote rank names.

\subsection*{Full Cascade}

\begin{align}
\triangleright~\text{Initializations:}\quad
  & H_{l,b_1,b_0,\,i:i=0,d,n} = 0 \\
\phase{Embeddings}
\LEX_{l:l=0,b_1,b_0,i,ed}
  &= \WET_{v,ed} \cdot \Tokens_{b_1,b_0,i,v}\notag\\
  &:: \bigwedge \leftarrow(\cap) \bigvee \mathrm{ANY}_v
\end{align}

\begin{align}
\triangleright~\text{Extended Einsum:}\;& \notag\\
\phase{Add + Norm}
\LEX_{l,b_1,b_0,i,ed}
  &= \EY_{l-1,b_1,b_0,i,ed} + \LEX_{l-1,b_1,b_0,i,ed} \\
\PX_{l,b_1,b_0,i,ed}
  &= \LEX_{l,b_1,b_0,i,ed} \cdot \LEX_{l,b_1,b_0,i,ed} \\
\NUM_{l,b_1,b_0,i}
  &= \PX_{l,b_1,b_0,i,ed} :: \bigvee +_{ed} \\
\MEX_{l,b_1,b_0,i}
  &= \tfrac{\NUM_{l,b_1,b_0,i}}{\ED} \\
\SQEX_{l,b_1,b_0,i}
  &= \tfrac{1}{\sqrt{\MEX_{l,b_1,b_0,i}+\epsilon}} \\
\NEX_{l,b_1,b_0,i,ed}
  &= \LEX_{l,b_1,b_0,i,ed}\cdot \SQEX_{l,b_1,b_0,i}\cdot \WEX_{ed}
\end{align}

\begin{align}
\phase{Project $x$}
\TTX_{l,b_1,b_0,i,d}
  &= \WTTX_{l,ed,d}\cdot \NEX_{l,b_1,b_0,i,ed} \\
\RX_{l,b_1,b_0,i,d}
  &= \WRX_{l,ed,d}\cdot \NEX_{l,b_1,b_0,i,ed} \\
\phase{Conv1d($x$)}
\TX_{l,b_1,b_0,i,d}
  &= \WConv_{l,d,f}\cdot \TTX_{l,b_1,b_0,i-f,d}+BTX_{l,d} \\
\phase{SiLU($x$)}
X_{l,b_1,b_0,i,d}
  &= \TX_{l,b_1,b_0,i,d}\times \sigma\!\big(\TTX_{l,b_1,b_0,i,d}\big)
\end{align}

\begin{align}
\phase{SSM time!}
\TT\Delta_{l,b_1,b_0,i,r}
  &= \WTT\Delta_{l,d,r}\cdot X_{l,b_1,b_0,i,d} \\
B_{l,b_1,b_0,i,n}
  &= \WB_{l,d,n}\cdot X_{l,b_1,b_0,i,d} \\
C_{l,b_1,b_0,i,n}
  &= \WC_{l,d,n}\cdot X_{l,b_1,b_0,i,d} \\
T\Delta_{l,b_1,b_0,i,d}
  &= (\WT\Delta_{l,r,d}\cdot \TT\Delta_{l,b_1,b_0,i,r})_{l,b_1,b_0,i,d} + B\Delta_d \\
\phase{Softplus to get $\Delta$}
\Delta_{l,b_1,b_0,i,d}
  &= \tfrac{1}{\beta}\log\!\Big(1_{l,b_1,b_0,i,d}+\exp(\beta \TT\Delta_{l,b_1,b_0,i,d})\Big)
\end{align}

\begin{align}
\phase{Selective scan (Euler for $B$)}
\bar{A}_{l,b_1,b_0,i,d,n}
  &= \exp\!\big(\Delta_{l,b_1,b_0,i,d}\,A_{l,d,n}\big) \\
\bar{B}_{l,b_1,b_0,i,d,n}
  &= \Delta_{l,b_1,b_0,i,d}\,B_{l,b_1,b_0,i,n} \\
\HX_{l,b_1,b_0,i,d,n}
  &= \bar{B}_{l,b_1,b_0,i,d,n}\,X_{l,b_1,b_0,i,d} \\
\HH_{l,b_1,b_0,i,d,n}
  &= \bar{A}_{l,b_1,b_0,i,d,n}\,H_{l,b_1,b_0,i-1,d,n} \\
H_{l,b_1,b_0,i,d,n}
  &= \HX_{l,b_1,b_0,i,d,n}+\HH_{l,b_1,b_0,i,d,n} \\
\SSIXY_{l,b_1,b_0,i,d}
  &= C_{l,b_1,b_0,i,n}\,H_{l,b_1,b_0,i,d,n} \\
\LY_{l,b_1,b_0,i,d}
  &= \SSIXY_{l,b_1,b_0,i,d}+D_d\,X_{l,b_1,b_0,i,d} \\
\phase{Mix in part of original input using SiLU}
Y_{l,b_1,b_0,i,d}
  &= \notag\\
  &\LY_{l,b_1,b_0,i,d}\,\Big(\RX_{l,b_1,b_0,i,d}\,\sigma(\RX_{l,b_1,b_0,i,d})\Big) \\
\phase{Project back to embedding space}
\EY_{l,b_1,b_0,i,ed}
  &= \WEY_{l,d,ed}\,Y_{l,b_1,b_0,i,d} \\
\diamond:\; l
  &\equiv L \\
\phase{LM only: Convert to logits (weight tying)}
\mathrm{logits}_{b_1,b_0,i,v}
  &= \WET_{ed,v}\cdot \NY_{b_1,b_0,i,ed}
\end{align}

\clearpage
\section{Mamba-2 Cascade}

\subsection{Rank Names}

\subsubsection*{Word/token length ranks}
\begin{itemize}
  \item $I$ --- sequence length
  \begin{itemize}
    \item $I_0$ --- sequence length of the prompt
  \end{itemize}
\end{itemize}

\subsubsection*{Vocabulary/embedding ranks}
\begin{itemize}[leftmargin=1.5em]
  \item $V$ --- dictionary size (words in vocabulary)
  \item $\ED$ --- input global space embedding
  \item $D$ --- channel, expanded from $\ED$ \quad ($D = E \times \ED = P \times Q$)
\end{itemize}

\subsubsection*{Head ranks}
\begin{itemize}[leftmargin=1.5em]
  \item $P$ --- number of SSM heads (the $A$ and $\Delta$ tensors act per head)
  \item $Q$ --- head dimension (channels within a single head), so that $D = P \times Q$
\end{itemize}

\subsubsection*{Other Ranks}
\begin{itemize}[leftmargin=1.5em]
  \item $N$ --- hidden state/memory dimension. The larger this is, the better the hidden state tensor is able to remember context
  \item $E$ --- expansion factor from $\ED$ to $D$
  \item $F$ --- window size for the causal convolution (correlation)
  \item $B$ --- total number of prompts (batch)
  \item $L$ --- number of layers
\end{itemize}

\subsection*{Input}

\subsubsection*{Input tensors}
\begin{itemize}[leftmargin=1.5em]
  \item \textcolor{myblue}{\(\mathrm{Tokens}^{I,V}\)} --- raw input (input to first layer only)
  \item \(\LEX^{I,ED}\) --- input after embedding (input to subsequent layers)
\end{itemize}

\subsection*{Principal Intermediate Tensors}
\begin{itemize}[leftmargin=1.5em]
  \item \(\Delta^{I,P}\) --- time step for discretization (one per head)
  \item \(\NEX^{I,ED}\) --- normalized input
  \item \(X^{I,P,Q}\) --- input to state-space model
  \item \(\bar{A}^{I,P}\) --- discretized \(A\) tensor (one per head)
  \item \(\bar{B}^{I,P,N}\) --- discretized \(B\) tensor
  \item \(H^{I,P,Q,N}\) --- hidden state/memory tensor
  \item \(\LY^{I,P,Q}\) --- state-space model output before gating
  \item \(\LLY^{I,P,Q}\) --- gated output before the output LayerNorm
\end{itemize}

\subsection*{Weight and Trained Tensors}
\begin{itemize}[leftmargin=1.5em]
  \item \textcolor{mygreen}{\(\WEX^{ED}\)} --- weight tensor to create \(\NEX\)
  \item \textcolor{mygreen}{\(\WTX^{ED,P,Q}\)} --- in-projection weight tensor to create \(\TX\)
  \item \textcolor{mygreen}{\(\WRX^{ED,P,Q}\)} --- in-projection weight tensor to create \(\RX\) (gate)
  \item \textcolor{mygreen}{\(\WTB^{ED,N}\)} --- in-projection weight tensor to create \(\TB\)
  \item \textcolor{mygreen}{\(\WTC^{ED,N}\)} --- in-projection weight tensor to create \(\TC\)
  \item \textcolor{mygreen}{\(\WT\Delta^{ED,P}\)} --- in-projection weight tensor to create \(T\Delta\) (for \(\Delta\))
  \item \textcolor{mygreen}{\(\WTTX^{P,Q,F}\)} --- convolution filter to create \(\TTX\)
  \item \textcolor{mygreen}{\(\WTTB^{N,F}\)} --- convolution filter to create \(\TTB\)
  \item \textcolor{mygreen}{\(\WTTC^{N,F}\)} --- convolution filter to create \(\TTC\)
  \item \textcolor{mygreen}{\(A^{P}\)} --- trained state-transition tensor (scalar per head)
  \item \textcolor{mygreen}{\(D^{P}\)} --- trained skip-connection tensor (scalar per head)
  \item \textcolor{mygreen}{\(\WNLLY^{P,Q}\)} --- weight tensor for the output LayerNorm (to create \(\NLLY\))
  \item \textcolor{mygreen}{\(\WEY^{P,Q,ED}\)} --- weight tensor to project back to the embedding space
  \item \textcolor{mygreen}{\(\WET^{ED,V}\)} --- weight tensor to project from the embedding space to logits (also projects from token space to embedding space)
\end{itemize}

\noindent\textit{Note:} Superscripts denote rank names.

\subsection*{Full Cascade}

\begin{align}
\triangleright~\text{Initializations:}\quad
  & H_{l,b,\,i:i=0,p,q,n} = 0 \\
\phase{Embeddings}
\LEX_{l:l=0,b,i,ed}
  &= \WET_{v,ed} \cdot \Tokens_{b,i,v}\notag\\
  &:: \bigwedge \leftarrow(\cap) \bigvee \mathrm{ANY}_v
\end{align}

\begin{align}
\triangleright~\text{Extended Einsum:}\;& \notag\\
\phase{Add + Norm}
\LEX_{l,b,i,ed}
  &= \EY_{l-1,b,i,ed} + \LEX_{l-1,b,i,ed} \\
\PX_{l,b,i,ed}
  &= \LEX_{l,b,i,ed} \cdot \LEX_{l,b,i,ed} \\
\NUM_{l,b,i}
  &= \PX_{l,b,i,ed} :: \bigvee +_{ed} \\
\MEX_{l,b,i}
  &= \tfrac{\NUM_{l,b,i}}{\ED} \\
\SQEX_{l,b,i}
  &= \tfrac{1}{\sqrt{\MEX_{l,b,i}+\epsilon}} \\
\NEX_{l,b,i,ed}
  &= \LEX_{l,b,i,ed}\cdot \SQEX_{l,b,i}\cdot \WEX_{ed}
\end{align}

\begin{align}
\phase{In-proj ($x$, $r$, $B$, $C$, $\Delta$)}
\TX_{l,b,i,p,q}
  &= \WTX_{l,ed,p,q}\cdot \NEX_{l,b,i,ed} \\
\RX_{l,b,i,p,q}
  &= \WRX_{l,ed,p,q}\cdot \NEX_{l,b,i,ed} \\
\TB_{l,b,i,n}
  &= \WTB_{l,ed,n}\cdot \NEX_{l,b,i,ed} \\
\TC_{l,b,i,n}
  &= \WTC_{l,ed,n}\cdot \NEX_{l,b,i,ed} \\
T\Delta_{l,b,i,p}
  &= \WT\Delta_{l,ed,p}\cdot \NEX_{l,b,i,ed} \\
\phase{Conv1d ($x$, $B$, $C$)}
\TTX_{l,b,i,p,q}
  &= \WTTX_{l,p,q,f}\cdot \TX_{l,b,i-f,p,q} \\
\TTB_{l,b,i,n}
  &= \WTTB_{l,n,f}\cdot \TB_{l,b,i-f,n} \\
\TTC_{l,b,i,n}
  &= \WTTC_{l,n,f}\cdot \TC_{l,b,i-f,n} \\
\phase{SiLU ($x$, $B$, $C$)}
X_{l,b,i,p,q}
  &= \TTX_{l,b,i,p,q}\times \sigma\!\big(\TTX_{l,b,i,p,q}\big) \\
B_{l,b,i,n}
  &= \TTB_{l,b,i,n}\times \sigma\!\big(\TTB_{l,b,i,n}\big) \\
C_{l,b,i,n}
  &= \TTC_{l,b,i,n}\times \sigma\!\big(\TTC_{l,b,i,n}\big)
\end{align}

\begin{align}
\phase{Softplus to get $\Delta$}
\Delta_{l,b,i,p}
  &= \tfrac{1}{\beta}\log\!\Big(1_{l,b,i,p}+\exp(\beta\, T\Delta_{l,b,i,p})\Big) \\
\phase{Main recurrence ($A$ is a per-head scalar)}
\bar{A}_{l,b,i,p}
  &= \exp\!\big(\Delta_{l,b,i,p}\,A_{l,p}\big) \\
\bar{B}_{l,b,i,p,n}
  &= \Delta_{l,b,i,p}\,B_{l,b,i,n} \\
\HX_{l,b,i,p,q,n}
  &= \bar{B}_{l,b,i,p,n}\,X_{l,b,i,p,q} \\
\HH_{l,b,i,p,q,n}
  &= \bar{A}_{l,b,i,p}\,H_{l,b,i-1,p,q,n} \\
H_{l,b,i,p,q,n}
  &= \HX_{l,b,i,p,q,n}+\HH_{l,b,i,p,q,n} \\
\phase{SSM output}
\SSIXY_{l,b,i,p,q}
  &= C_{l,b,i,n}\,H_{l,b,i,p,q,n} \\
\LY_{l,b,i,p,q}
  &= \SSIXY_{l,b,i,p,q}+D_p\,X_{l,b,i,p,q}
\end{align}

\begin{align}
\phase{Gate with SiLU($\RX$)}
\LLY_{l,b,i,p,q}
  &= \notag\\
  &\LY_{l,b,i,p,q}\,\Big(\RX_{l,b,i,p,q}\,\sigma(\RX_{l,b,i,p,q})\Big) \\
\phase{Output LayerNorm (RMS over $D = P\times Q$)}
\PLLY_{l,b,i,p,q}
  &= \LLY_{l,b,i,p,q}\cdot \LLY_{l,b,i,p,q} \\
\PLLYNUM_{l,b,i}
  &= \PLLY_{l,b,i,p,q} :: \bigvee +_{p,q} \\
\MLLY_{l,b,i}
  &= \tfrac{\PLLYNUM_{l,b,i}}{D} \\
\SQLLY_{l,b,i}
  &= \tfrac{1}{\sqrt{\MLLY_{l,b,i}+\epsilon}} \\
\NLLY_{l,b,i,p,q}
  &= \LY_{l,b,i,p,q}\cdot \SQLLY_{l,b,i}\cdot \WNLLY_{p,q} \\
\phase{Project back to embedding space}
\EY_{l,b,i,ed}
  &= \WEY_{l,p,q,ed}\,\NLLY_{l,b,i,p,q} \\
\diamond:\; l
  &\equiv L \\
\phase{LM only: Convert to logits (weight tying)}
\mathrm{logits}_{i,b,v}
  &= \WET_{ed,v}\,\EY_{l:l=L-1,b,i,ed} \\
\phase{Retrieve output token}
\Tokens_{b,i+1,v}
  &= \mathrm{Select1}\!\big(\mathrm{logits}_{i:i=I_0-1,b,v}\big)
\end{align}